\documentclass[aip,rsi,reprint]{revtex4-1}
\usepackage[pdftex]{graphicx}% Include figure files / force pdf mode %---
\pdfoutput=1
\usepackage{hyperref} 
\usepackage{dcolumn}% Align table columns on decimal point

\draft % marks overfull lines with a black rule on the right
\begin{document}

%\preprint{}

\title{The Cryogenic Storage Ring CSR} %Title of paper



% ------------------------------------
%\author{}

\author{R.~\surname{von~Hahn}}\affiliation{Max-Planck-Institut f\"ur Kernphysik, 69117 Heidelberg, Germany}%
\author{A.~Becker}\affiliation{Max-Planck-Institut f\"ur Kernphysik, 69117 Heidelberg, Germany}%
\author{F.~Berg}\altaffiliation[Present address: ]{Paul Scherrer Institut (PSI), 5232 Villigen, Switzerland}\affiliation{Max-Planck-Institut f\"ur Kernphysik, 69117 Heidelberg, Germany}%
\author{K.~Blaum}\affiliation{Max-Planck-Institut f\"ur Kernphysik, 69117 Heidelberg, Germany}%
\author{C.~Breitenfeldt}\affiliation{Max-Planck-Institut f\"ur Kernphysik, 69117 Heidelberg, Germany}\affiliation{Institut f\"ur Physik, Ernst-Moritz-Arndt-Universit\"at, 17487 Greifswald, Germany}%
\author{H.~Fadil}\affiliation{Max-Planck-Institut f\"ur Kernphysik, 69117 Heidelberg, Germany}%
\author{F.~Fellenberger}\affiliation{Max-Planck-Institut f\"ur Kernphysik, 69117 Heidelberg, Germany}%
\author{M.~Froese}\affiliation{Max-Planck-Institut f\"ur Kernphysik, 69117 Heidelberg, Germany}%
\author{S.~George}\affiliation{Max-Planck-Institut f\"ur Kernphysik, 69117 Heidelberg, Germany}%
\author{J.~G\"ock}\affiliation{Max-Planck-Institut f\"ur Kernphysik, 69117 Heidelberg, Germany}%
\author{M.~Grieser}\affiliation{Max-Planck-Institut f\"ur Kernphysik, 69117 Heidelberg, Germany}%
\author{F.~Grussie}\affiliation{Max-Planck-Institut f\"ur Kernphysik, 69117 Heidelberg, Germany}%
\author{E.~A.~Guerin}\affiliation{Max-Planck-Institut f\"ur Kernphysik, 69117 Heidelberg, Germany}%
\author{O.~Heber}\affiliation{Weizmann Institute of Science, Rehovot 76100, Israel}%
\author{P.~Herwig}\affiliation{Max-Planck-Institut f\"ur Kernphysik, 69117 Heidelberg, Germany}%
\author{J.~Karthein}\affiliation{Max-Planck-Institut f\"ur Kernphysik, 69117 Heidelberg, Germany}%
\author{C.~Krantz}\affiliation{Max-Planck-Institut f\"ur Kernphysik, 69117 Heidelberg, Germany}%
\author{H.~Kreckel}\affiliation{Max-Planck-Institut f\"ur Kernphysik, 69117 Heidelberg, Germany}%
\author{M.~Lange}\affiliation{Max-Planck-Institut f\"ur Kernphysik, 69117 Heidelberg, Germany}%
\author{F.~Laux}\affiliation{Max-Planck-Institut f\"ur Kernphysik, 69117 Heidelberg, Germany}%
\author{S.~Lohmann}\affiliation{Max-Planck-Institut f\"ur Kernphysik, 69117 Heidelberg, Germany}%
\author{S.~Menk}\altaffiliation[Present address: ]{RIKEN, Wako-shi, Saitama 351-0198, Japan}\affiliation{Max-Planck-Institut f\"ur Kernphysik, 69117 Heidelberg, Germany}%
\author{C.~Meyer}\affiliation{Max-Planck-Institut f\"ur Kernphysik, 69117 Heidelberg, Germany}%
\author{P.~M.~Mishra}\affiliation{Max-Planck-Institut f\"ur Kernphysik, 69117 Heidelberg, Germany}%
\author{O.~Novotn\'y}\affiliation{Max-Planck-Institut f\"ur Kernphysik, 69117 Heidelberg, Germany}\affiliation{Columbia Astrophysics Laboratory, Columbia University, New York, \mbox{New York 10027, USA}}%
\author{A.~P.~\surname{O'Connor}}\affiliation{Max-Planck-Institut f\"ur Kernphysik, 69117 Heidelberg, Germany}%
\author{D.~A.~Orlov}\affiliation{Max-Planck-Institut f\"ur Kernphysik, 69117 Heidelberg, Germany}%
\author{M.~L.~Rappaport}\affiliation{Weizmann Institute of Science, Rehovot 76100, Israel}%
\author{R.~Repnow}\affiliation{Max-Planck-Institut f\"ur Kernphysik, 69117 Heidelberg, Germany}%
\author{S.~Saurabh}\affiliation{Max-Planck-Institut f\"ur Kernphysik, 69117 Heidelberg, Germany}%
\author{S.~Schippers}\affiliation{I.~Physikalisches Institut, Justus-Liebig-Universit\"at Gie\ss{}en, 35392 Gie\ss{}en, Germany}%
\author{C.~D.~Schr\"oter}\affiliation{Max-Planck-Institut f\"ur Kernphysik, 69117 Heidelberg, Germany}%
\author{D.~Schwalm}\affiliation{Max-Planck-Institut f\"ur Kernphysik, 69117 Heidelberg, Germany}\affiliation{Weizmann Institute of Science, Rehovot 76100, Israel}%
\author{L.~Schweikhard}\affiliation{Institut f\"ur Physik, Ernst-Moritz-Arndt-Universit\"at, 17487 Greifswald, Germany}%
\author{T.~Sieber}\altaffiliation[Present address: ]{Helmholtzzentrum f\"ur Schwerionenforschung (GSI), 64291 Darmstadt, Germany}\affiliation{Max-Planck-Institut f\"ur Kernphysik, 69117 Heidelberg, Germany}%
\author{A.~Shornikov}\altaffiliation[Present address: ]{CERN, 1211 Gen\`eve, Switzerland}\affiliation{Max-Planck-Institut f\"ur Kernphysik, 69117 Heidelberg, Germany}%
\author{K.~Spruck}\affiliation{I.~Physikalisches Institut, Justus-Liebig-Universit\"at Gie\ss{}en, 35392 Gie\ss{}en, Germany}\affiliation{Max-Planck-Institut f\"ur Kernphysik, 69117 Heidelberg, Germany}%
\author{S.~\surname{Sunil Kumar}}\affiliation{Max-Planck-Institut f\"ur Kernphysik, 69117 Heidelberg, Germany}%
\author{J.~Ullrich}\altaffiliation[Present address: ]{Physikalisch-Technische Bundesanstalt (PTB), 38116 Braunschweig, Germany}\affiliation{Max-Planck-Institut f\"ur Kernphysik, 69117 Heidelberg, Germany}%
\author{X.~Urbain}\affiliation{Institute of Condensed Matter and Nanosciences, Universit\'e Catholique de Louvain, 1348 Louvain-la-Neuve, Belgium}%
\author{S.~Vogel}\affiliation{Max-Planck-Institut f\"ur Kernphysik, 69117 Heidelberg, Germany}%
\author{P.~Wilhelm}\affiliation{Max-Planck-Institut f\"ur Kernphysik, 69117 Heidelberg, Germany}%
\author{A.~Wolf}\email[Electronic Mail: ]{A.Wolf@mpi-hd.mpg.de}\affiliation{Max-Planck-Institut f\"ur Kernphysik, 69117 Heidelberg, Germany}%
\author{D.~Zajfman}\affiliation{Weizmann Institute of Science, Rehovot 76100, Israel}

%\date{24 May 2016}

\hyphenation{cryo-pumping}
\hyphenation{cryo-condens-ation}
\hyphenation{Karls-ruhe}

\begin{abstract}
  An electrostatic cryogenic storage ring, CSR, for beams of anions and cations with up to 300~keV kinetic energy per unit charge has
  been designed, constructed and put into operation.  With a circumference of 35~m, the ion-beam vacuum chambers and all beam optics
  are in a cryostat and cooled by a closed-cycle liquid helium system.  At temperatures as low as $(5.5\pm1)$~K inside the ring,
  storage time constants of several minutes up to almost an hour were observed for atomic and molecular, anion and cation beams at an
  energy of 60~keV.  The ion-beam intensity, energy-dependent closed-orbit shifts (dispersion) and the focusing properties of the
  machine were studied by a system of capacitive pickups.  The Schottky-noise spectrum of the stored ions revealed a broadening of the
  momentum distribution on a time scale of 1000~s.  Photodetachment of stored anions was used in the beam lifetime measurements.  The
  detachment rate by anion collisions with residual-gas molecules was found to be extremely low.  A residual-gas density below
  140~cm$^{-3}$ is derived, equivalent to a room-temperature pressure below 10$^{-14}$ mbar.  Fast atomic, molecular and cluster ion
  beams stored for long periods of time in a cryogenic environment will allow experiments on collision- and radiation-induced
  fragmentation processes of ions in known internal quantum states with merged and crossed photon and particle beams.
\end{abstract}


\onecolumngrid
\vspace*{-2.5\baselineskip}
\hspace*{\fill}Manuscript of an article accepted by Review of Scientific Instruments\\
\hspace*{\fill}(Dated: 24 May 2016)
\vspace{-0.5\baselineskip}
\twocolumngrid
\maketitle %\maketitle must follow title, authors, abstract and \pacs

%  \newlength{\bskip}\setlength{\bskip}{-3mm}

\section{Introduction}

Ion storage rings have proven to be unique tools for investigating properties and interactions of atomic and molecular ions.  Mass-selected fast ion beams are used for studying collisions with well-defined energy in merged beams \cite{schuch_atomic_2007,larsson_dissociative_1997,schippers_electronion_2015} or in stationary in-ring targets \cite{stoehlker_lamb_2000,schmidt_evidence_2008,schneider_2013}.  Considering its astrophysical relevance, the dissociative recombination of molecular ions with electrons is being intensely investigated \cite{larsson_dissociative_1997}, including fragmentation into multiple neutral products for polyatomic species \cite{geppert_dissociative_2008}.  Another field of study is the de-excitation of internal states in atomic ions \cite{schmidt_high-precision_1994} and in small \cite{amitay_dissociative_1998} to very large \cite{andersen_radiative_1996} molecular systems.

To date, many of these studies were carried out in magnetic storage rings \cite{wolf_heavy-ion_1999} with particle kinetic energies in the MeV range and about 50 to 110~m ring circumference.  As a complementary technique, electrostatic storage rings for atomic and molecular physics were pioneered with the ELISA facility \cite{moller_elisa_1997} in Aarhus, with an ion-orbit circumference of 7~m.  Even though the accessible beam energies are more limited with electrostatic bending, electrostatic optics are simpler to use as magnetic hysteresis effects are absent and fringe fields are smaller.  Most importantly, however, electrostatic rings do not restrict the mass-to-charge ratio of the particles at a given energy and ions of heavy atoms and molecules, even clusters and large biomolecules, become available for experimental studies in such devices.  Thus, electrostatic operation greatly widens the scope \cite{andersen_physics_2004} of atomic and molecular physics in ion storage rings.  Recent experiments even extended to electron interactions with complex molecular ions \cite{tanabe_electrostatic_2004}, included partial cooling of the storage-ring environment to liquid nitrogen temperatures \cite{jinno_tmu_2004} and introduced electrostatic storage rings only 1~m on a side \cite{pedersen_characterization_2015} and beyond (down to $\sim$\,0.3~m ion-orbit circumference \cite{bernard_tabletop_2008,martin_fast_2013}).  Cryogenically cooling the complete structure of electrostatic ion storage rings was realized for the double storage ring DESIREE \cite{rensfelt_desiree_2004,thomas_double_2011} of $2\times8.8$~m ion-orbit circumference.  Successful operation of this ring, with the particular aim of anion--cation merged beams experiments, was recently reported \cite{schmidt_first_2013,backstrom_storing_2015}.  Another cryogenic, $\sim$\,3~m circumference single-ring facility \cite{nakano_cryogenic_2012} aiming at merged beams studies with lasers and particle beams is under commissioning.
 
The electrostatic cryogenic ion storage ring, CSR, at the Max Planck Institute for Nuclear Physics in Heidelberg, Germany, was proposed \cite{zajfman_2005} for collisional and laser-interaction studies over long storage times with fast atomic, molecular and cluster ion beams.  The cryogenic facility offers an ambient temperature of $\sim$\,6~K for the stored ion beam and extended field-free straight sections for in-ring experiments, often involving the detection of fast products released by reactive collisions of the stored ions.  The layout of the CSR also permits the addition of a collinear merged electron beam for electron collision studies and for applying phase-space cooling to a wide range of stored ion species.

Here we report on the development, construction and first operation of this cryogenic electrostatic storage ring.  Based on the conceptual goals, we discuss the layout of the system, highlighting the ion-optical, cryogenic and mechanical design of the CSR.  The completed ring was cooled to cryogenic temperature and subsequently operated successfully storing ion beams of kinetic energies in the region of 60~keV.  The stored beams were probed both by capacitive pickups and by laser interaction.  We describe the first generation of experimental devices installed in the ring and provide a set of measurements characterizing the storage of ion beams for durations of many minutes up to a few hours.  Information is provided on the beam optics and the energy distribution of the stored ions and on the extremely low residual-gas densities reached by the cryogenic pumping system.

\section{The CSR concept}
\label{concept}

Cryogenic storage rings offer decisive advantages.  Firstly, one can obtain very long storage times for fast beams with multi-keV kinetic energies of ions in a wide mass range, making use of extremely rarified background gas.  Secondly, keeping the vacuum vessel and all beam line elements in the line-of-sight of the stored ions at very low temperature suppresses radiative (blackbody) excitation of the stored ions.  Together, this opens a wealth of entirely new opportunities for collisional, radiative and optical studies of internally cold, fast ions in an extended range of masses and molecular complexity.

\begin{figure}[t]
    \centering
    \includegraphics[width=86mm]{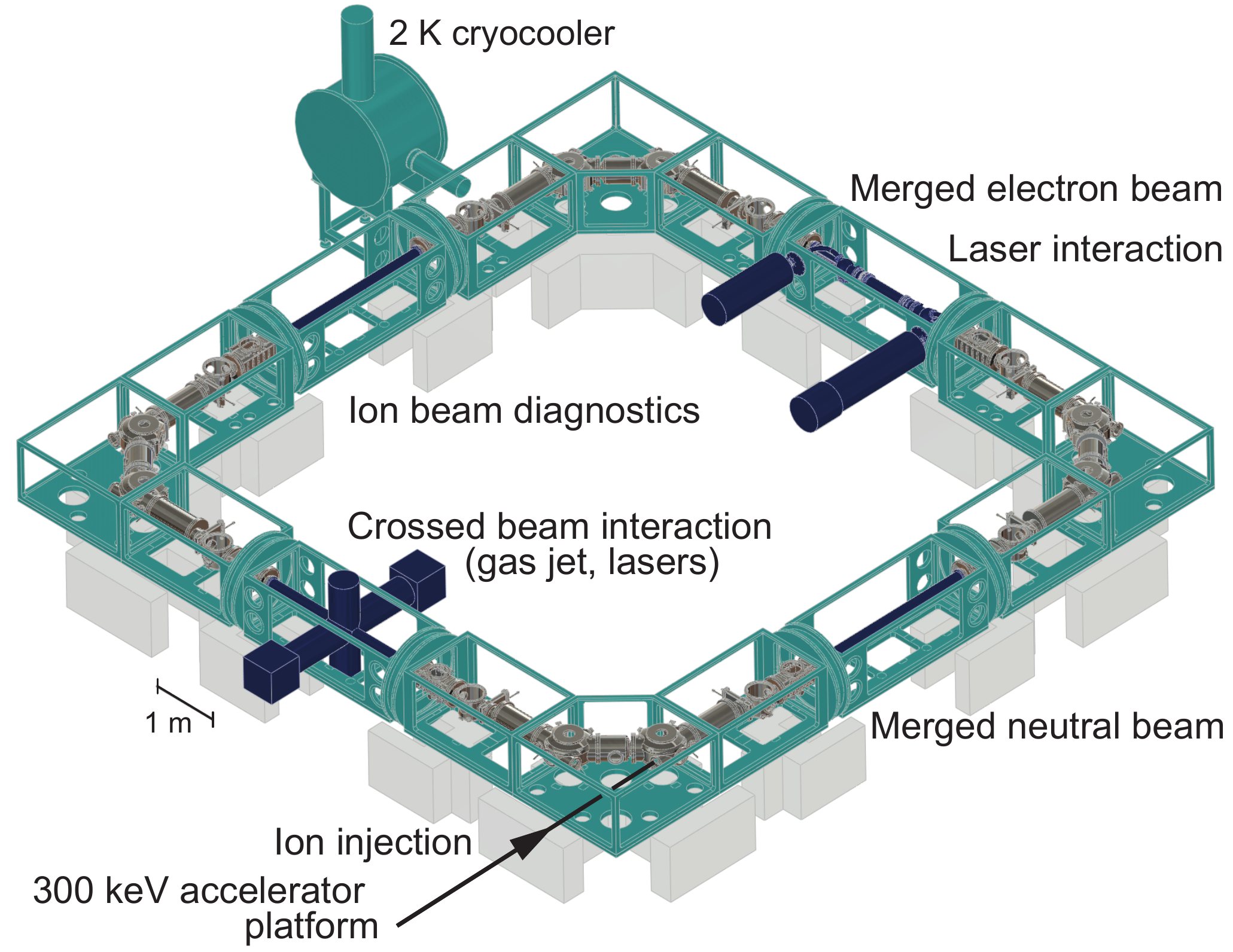}
    \caption{Overview of the cryogenic storage ring CSR showing the cryostat structure and the main inner cryogenic vacuum chambers (see Sec.~\ref{mechanical}) together with the principal experimental regions realized or in project.}
    \label{fig:CSRoverview}
\end{figure}

The CSR (Fig.~\ref{fig:CSRoverview}) is designed to accommodate long straight sections for in-ring experiments, kept free of ion-optical deflecting fields.  One of the straight sections will house a collinear electron interaction region, where the ions will merge over a length of $\sim$\,1~m with a near-monoenergetic beam of electrons.  Merged-beams experiments offer particularly favorable conditions when the average longitudinal velocities of both beams can be matched.  In order to achieve this for as wide a range of stored ion masses as possible, a high kinetic energy of the ion beam is desirable.  Under the condition of matched beam velocities one can envisage performing translational phase-space cooling of the stored ion beam, using the method of electron cooling \cite{poth_electron_1990} established earlier at low-energy magnetic storage rings.  Moreover, the center-of-mass electron--ion collision energies in the interaction region, limited predominantly by the velocity spread of the electron beam, can be as low as $\sim$\,1~meV.  A photocathode electron-beam source \cite{orlov_long_2009}, previously operated at the magnetic storage ring TSR, is being adapted for use in the CSR.  The electron-beam device is combined with particle detectors that can observe fast-moving products from collision-induced fragmentation reactions in the merged-beams interaction region.  To meet all these goals, the basic layout of CSR is characterized by electrostatic deflecting and focusing cells for a beam energy of 300~keV for singly-charged ions, arranged in a square geometry providing four main straight sections, devoid of beam-optical fields, of 2.6~m length each.

In addition to the merged electron-beam device for electron cooling and electron--ion collision measurements, the CSR has further options for collisional and radiative studies.  Laser beams can be crossed or merged with the circulating ions.  Another straight section will be designated for investigating collisional interactions of the stored ion beam with neutral atoms in a merged and velocity-matched beam.  For the third straight section, an integrated ``reaction-microscope'' setup \cite{ullrich_recoil-ion_2003} is designed that will serve as a powerful detector for electrons and heavy fragments emitted from a crossed-beam interaction region (see Fig.\ \ref{fig:CSRoverview}).  It is foreseen to cross the stored ion beam in this region with a thin neutral molecular beam (gas-jet target) as well as with laser pulses.  Finally, the fourth straight section is used for diagnostic elements applying, in particular, capacitive pickup electrodes to non-destructively observe the stored ion beam.

The CSR facility also comprises a 300~kV accelerator platform suitable for interfacing with a wide range of ion-chemistry and plasma-type sources for cations and anions as well as highly charged ions.  After electrostatic acceleration, the ion beam is deflected and mass analyzed by two large 45$^\circ$ bending magnets with 1.3~m orbit radius before being transported into the injection straight section of the CSR through a $\sim$\,10~m long beam line via electrostatic quadrupole elements.  In a later upgrade, the magnets of the injection line can be bypassed by a 90$^\circ$ electrostatic deflector.  This will allow injection into the CSR even for very heavy ions which, at their correspondingly small velocities, can be mass-analyzed by time-of-flight filtering.

The straight section of the injection beam line will also be equipped for producing a beam of fast neutral atoms that, further downstream, will overlap the ions stored in the ring.  For this purpose, a separately produced beam of anions will be merged into the injection line, during which time an ion beam is stored in the CSR.  The elements of the injection beam line will be switched to the appropriate focusing of the anion beam that will generate a fast moving neutral beam in an efficient photodetachment region \cite{oconnor_generation_2015} before entering the CSR.

Thus, with these experimental setups, the CSR facility offers attractive options for conducting unprecedented studies on collision- and radiation-induced fragmentation processes of energetic atomic, molecular and cluster ion beams.  The cryogenic operation, in particular, makes it possible to define the internal quantum states of stored molecular ions prior to use in fast-beam experiments.

\section{Storage-ring layout}
\subsection{Outline and ion optical lattice}
\label{lattice}

\begin{figure}[t]
    \centering
    \includegraphics[width=86mm]{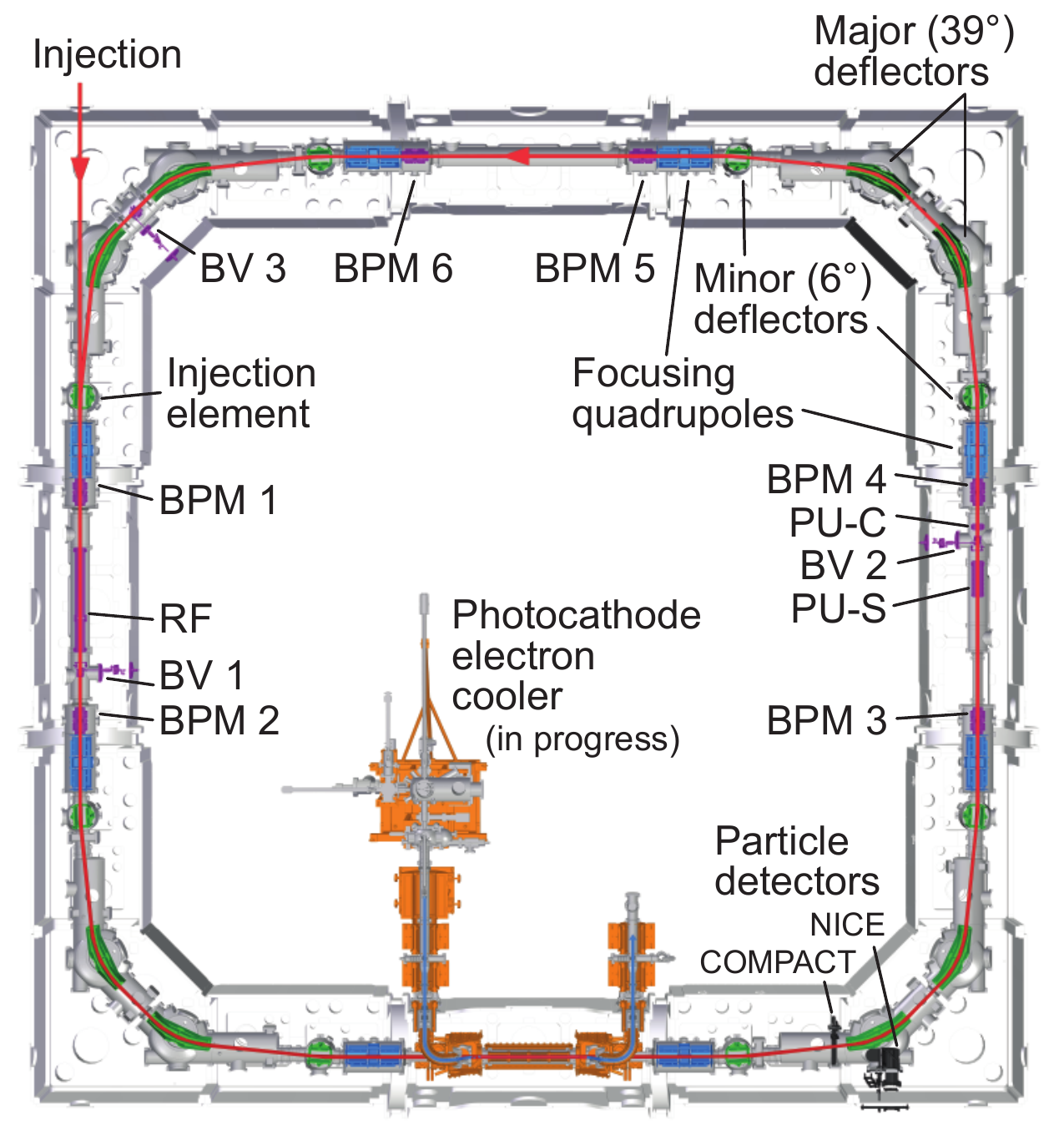}
    \caption{Components and layout of the CSR.  The circumference of the ion orbit is 35~m.  The main components of the ion optical lattice are marked in the upper right quadrant.  Further elements labeled by acronyms are the beam viewers BV, the beam-position monitors BPM, the current-pickup PU-C, the Schottky-noise pickup PU-S, the radiofrequency system for ion-beam bunching (RF) and the particle detectors discussed in Sec.~\ref{detectors}.}
    \label{fig:CSRlayout}
\end{figure}

\begin{figure}
    \centering
    \includegraphics[width=86mm]{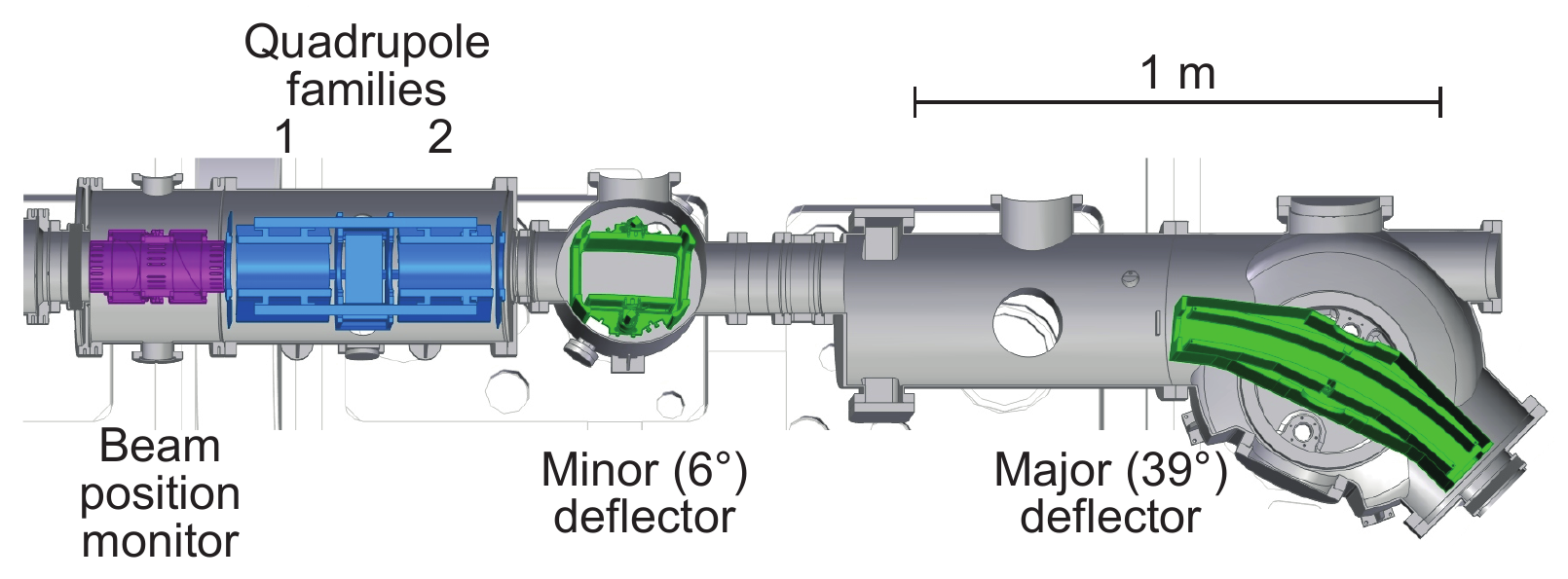}
 \caption{Enlarged view of a deflection and focusing cell of the CSR with the assignment of the quadrupole families (see text) and a beam position monitor.}
    \label{fig:CSRzoom}
\end{figure}

\begin{figure}
    \centering
    \includegraphics[width=82mm]{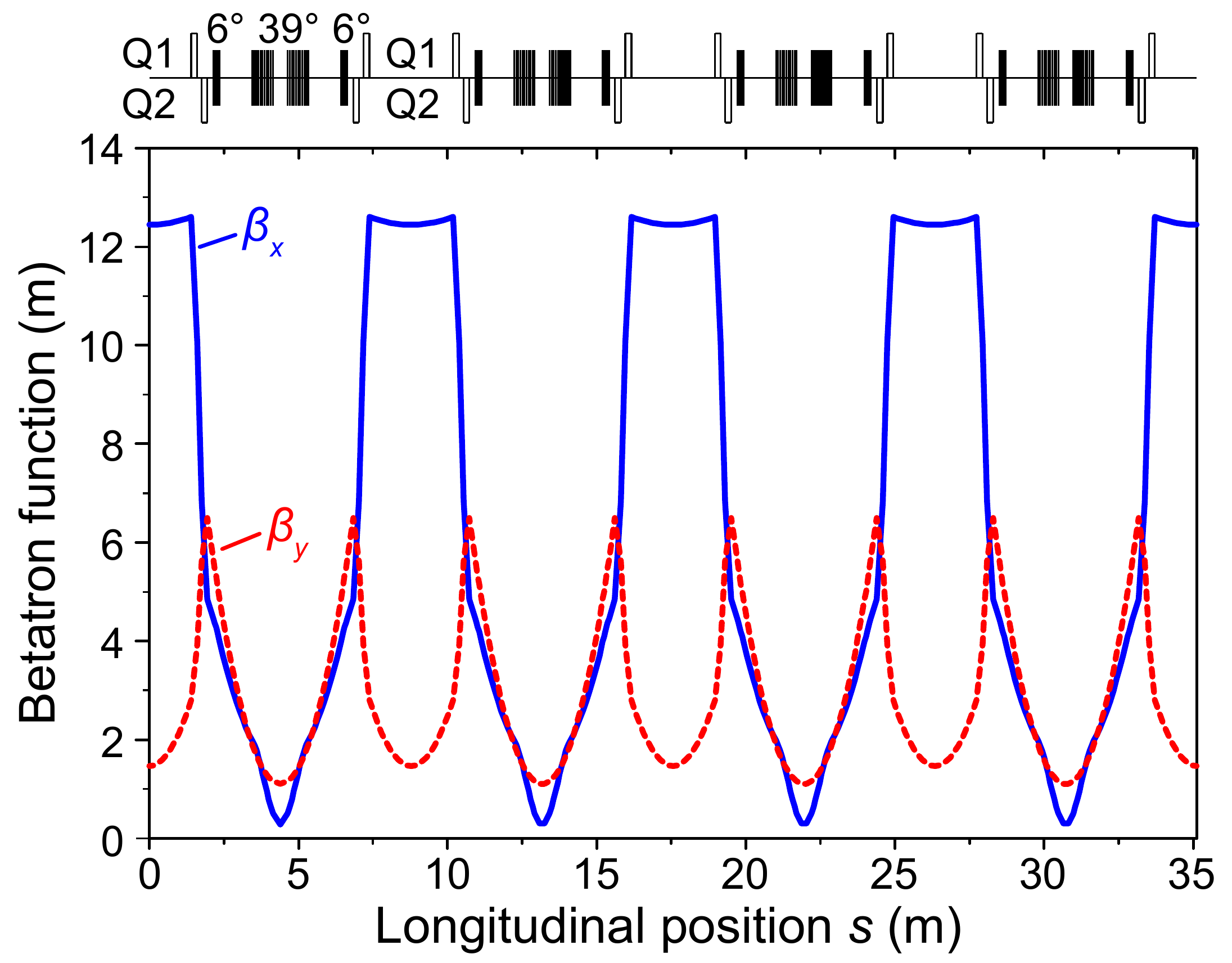}
    \caption{Horizontal and vertical betatron functions $\beta_{x}$ and  $\beta_{y}$, respectively, calculated by the MAD8 \cite{MAD8} code (lines as labeled) for the standard settings of the CSR ($Q_x=Q_y=2.59$).  The scheme indicates the arrangement along the ring circumference of the deflectors (marked by their angles) and of the quadrupole lenses of both families (marked Q1 and Q2).  Multiplied by the emittances, the betatron functions describe how the squares of the horizontal and vertical ion beam sizes vary along the storage ring orbit (see the text). }
    \label{fig:beta}
\end{figure}

The components of the storage ring, with its four deflection quadrants and four straight sections, are shown in Fig.~\ref{fig:CSRlayout}.  The ion optical lattice is formed by electrostatic cylindrical deflectors and quadrupole doublets. The 90$^\circ$ bends at each corner are comprised of two 6$^\circ$ (minor) deflectors, with a bending radius of 2~m, and two 39$^\circ$ (major) deflectors with a bending radius of 1~m.  The nominal ring circumference resulting from this lattice is $L_0=35.12$~m.  This layout is advantageous in that neutral particles or a laser beam can be merged, with the stored ions along one of the straight sections, making use of the gap between the minor and the major deflectors.  Moreover, the geometry leaves a relatively large open angular cone ($\pm1^\circ$) for observing neutral reaction fragments, produced in the middle of a straight section, on a detector behind the major deflectors.  Similarly, it offers many options for detecting mass- and charge-analysed reaction fragments.  Another advantage is the ease of single-turn injection, in which only the relatively small 6$^\circ$ deflector needs to be switched at a rate fast in comparison to the revolution time (see Sec.~\ref{injection}).  Both the low capacitance of the 6$^\circ$ deflector and the reduced effect of inaccuracies in the voltage switching, due to the small deflection angle, contribute to this advantage.

To focus the beam, each of the four corners contains two quadrupole doublets, one placed before and one after the four deflectors.  These doublets have in total 16 quadrupole lens elements out of which all those lying further away from the minor deflectors are grouped into quadrupole family 1, while those lying next to the deflectors are grouped in family 2 (see Fig.\ \ref{fig:CSRzoom}).  Single voltages $U_{1}$ and $U_{2}$ are used, respectively, to control the focusing strength for all members of these families.  The standard settings of the quadrupole strengths correspond to betatron tunes \cite{courant_theory_1958} of $Q_x=Q_y=2.59$ in the horizontal ($x$) and vertical ($y$) directions.  The coordinates refer to an orthogonal beam reference frame with the longitudinal coordinate $s$ along the central orbit of the storage ring and the $x$ coordinate lying perpendicular to it in the bending plane.

The betatron functions $\beta_{x}(s)$ and $\beta_{y}(s)$ for this standard setting are shown in Fig.~\ref{fig:beta}.  The ion-beam size in each direction varies as $\sqrt{\beta_{x(y)}}$ and the beam divergence is essentially inversely proportional to the same function \cite{rossbach_schmueser_1994}.  Hence, the beams at the centers of the long straight sections are asymmetric, reaching their largest size and smallest divergence in the horizontal direction.  The betatron functions peak in the region of the quadrupoles and reach their minima
between the two 39$^\circ$ deflectors.  

For particles with a relative deviation $\Delta p/p_0$ from the nominal momentum $p_0$, the closed orbit is horizontally shifted from the nominal (central) orbit by $D_x(s)\Delta p/p_0$ according to the dispersion function \cite{rossbach_schmueser_1994} $D_x(s)$. For the standard lattice, $D_x$ reaches a maximum of 2.1~m in the middle of the straight sections.

\subsection{Optical elements}
%\label{}

The betatron functions of the CSR lattice, as described in Sec.~\ref{lattice}, were calculated using linear transfer matrices within the MAD8 program \cite{MAD8} under the assumption that all elements are hard-edged.  It is clear, however, that the actual optical elements exhibit deviations from the ideal field distributions, as well as fringing fields.  Hence, three-dimensional numerical calculations of the realistic CSR elements were performed using the finite-elements electrostatic code TOSCA \cite{Tosca}.  Exact geometrical models of the elements were created, and the solution space was discretized with a three-dimensional mesh. The fields, in particular the radial deflecting field $E_r$, are calculated for adequate boundary conditions, such as the grounded vacuum vessel, for suitable potentials on the electrodes and, wherever possible, accounting for symmetry conditions.  With numerical maps of the calculated fields, ions were then tracked through the storage ring using the respective module of TOSCA as well as the G4beamline code \cite{G4beamline}.  These more advanced calcuations are presented below and their results are compared to the MAD8 model.

\subsubsection{The major deflectors}

The main bending of the ion beam in each of the corners of the CSR is realized by two $39^\circ$ cylindrical deflectors.  Each of them consists of two cylindrical electrodes of 180~mm height and a gap of 60~mm.  The radius of curvature on the nominal orbit is 1~m.  For infinitely extended cylindrical geometry, voltages of $\pm17.74$~kV and $\mp18.28$~kV at the outer and inner electrodes of each deflector, respectively, would be required to bend ions of charge $\pm{}e$ with the nominal CSR energy of 300~keV on the zero equipotential.  Considering the finite transverse dimensions and the fringing fields, the voltages required to obtain the nominal deflection angle are computed numerically.

The calculated bending electric field in the middle of this element is found to be very similar to the ideal field of a cylindrical deflector.  Along the ion orbit, however, the fringing fields of the deflectors were found to have a strong influence on the particle dynamics.  To reduce this effect, grounded electrodes, each with a 15~mm gap in azimuthal direction, were installed as field clamps at both ends of each deflector.  The deflection angle of the $39^\circ$ deflector was precisely calculated by tracking a single 300~keV proton, starting from the center of the deflector and ending in a field free region.  We find that the proton is deflected by about $40^\circ$ and $41^\circ$ with and without grounded electrodes, respectively, which shows that the grounded clamping electrodes reduce the overall deflection through the smaller fringing field.

In the tracking calculations described below we found it of utmost importance that the deflection angle of this element is matched with the design value.  Therefore, the deflector (nominal arc) was trimmed (shortened) by 8.4~mm on each end to achieve the design deflection angle of $39^\circ$.

\subsubsection{The minor deflectors}

\begin{figure}
    \centering
    \includegraphics[width=80mm]{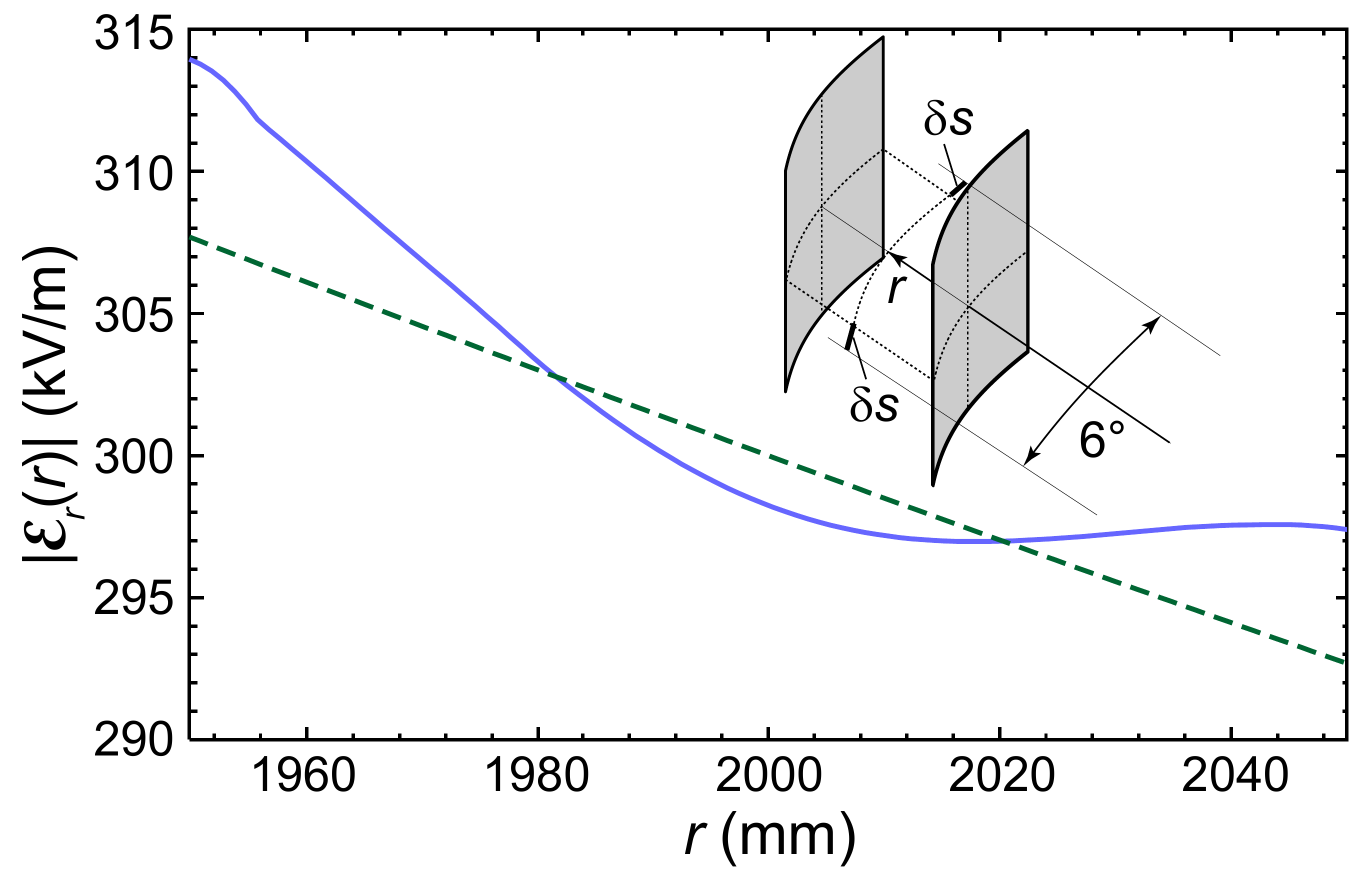}
    \caption{The magnitude $|{\cal E}_r|$ of the radial field inside a minor deflector as a function of the radial distance $r$ from the geometrical bending center (for 300~keV singly-charged ions, nominally $r=2000$~mm) along the intersection of the horizontal and vertical midplanes of the element (see the inset).  The ideal field for infinite electrode extension and 120~mm gap (dashed line) is compared to the numerical TOSCA calculation for the actual geometry (full line).  The inset also indicates the arc lengths $\delta s$ removed on each end for trimming the ion deflection angle (see the text).}
    \label{fig:6degree}
\end{figure}

\begin{figure}
    \centering\includegraphics[width=80mm]{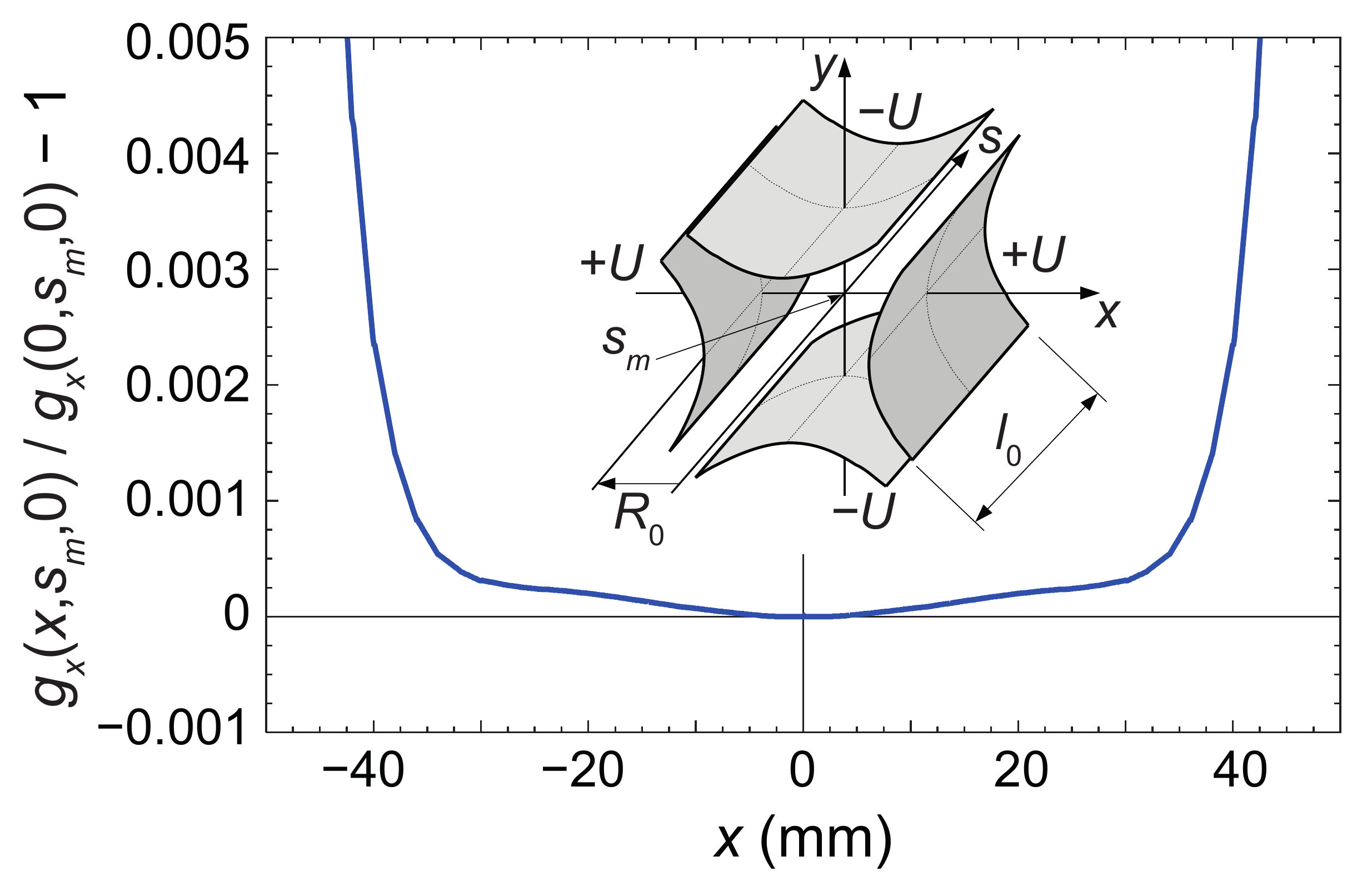}
    \caption{Relative deviation of the field gradient in the CSR focusing quadrupoles, $g_x(x,s_m,0)$, from its value $g_x(0,s_m,0)$ on the quadrupole axis as a function of the horizontal coordinate $x$.  The longitudinal coordinate lies in the middle of a quadrupole unit.  The inset indicates the basic quadrupole parameters discussed in the text.}
    \label{fig:Qgradr}
\end{figure}

The $6^\circ$ cylindrical deflectors of the CSR consist of two cylindrical electrodes with a height of 240~mm and a gap of 120~mm.  The ideal orbit has a radius of curvature of 2~m and the deflecting voltages for the ideal geometry are the same as for the major deflectors.  Similarly to the $39^\circ$ deflectors, grounded end-plates are utilized to clamp the fringing fields and the nominal arc is trimmed by $\delta s=9.1$~mm on each end in order to achieve a deflection angle of $6^\circ$.  The radial field as a function of the radial coordinate in the center of the deflector still deviates by several percent from the ideal field, as shown in Fig.~\ref{fig:6degree}.  This is mainly due to the limited azimuthal extent of the deflector and also to the rather large gap-to-height ratio of about $1:2$.  The precise field geometry of the  $6^\circ$ deflectors is taken into account in the tracking calculations described in Sec.~\ref{tracking}.

\subsubsection{The focusing quadrupoles}
\label{focquads}

For the focusing quadrupoles, hyperbolically shaped electrodes were chosen in order to achieve a quadrupolar field with minimal higher order components.  The inscribed radius of the quadrupole is $R_0=50$~mm and its pole length $l_0=200$~mm (see Fig.\ \ref{fig:Qgradr}).  A grounded shield is installed between the two units of each doublet.  Numerical calculations have shown that the two lenses of the doublet are thereby effectively decoupled.  

For realistic boundary conditions we have calculated the spatial distributions of the electric field gradients $g_x=\partial{\cal E}_x/\partial{x}$ and $g_y=\partial{\cal E}_y/\partial{y}$, where ${\cal E}_{x(y)}$ denote the components of the electric field.  In the quadrupole center, $s=s_m$, and in the horizontal and vertical mid-planes, respectively, the calculated gradients lie very close to the ideal values for infinite hyperbolic electrodes, $\hat{g}_x=\hat{g}$ and $\hat{g}_y=-\hat{g}$, with $\hat{g}=-2U/R_0^2$ and the definitions of Fig.~\ref{fig:Qgradr}.  The figure also shows the relative deviation of the gradient $g_x(x,s_m,0)$ from its value in the quadrupole center for a horizontal displacement $x$ starting at the mid-point $s=s_m$, $y=0$.  The relative deviation remains as small as $<\,2\times10^{-3}$ for $|x|<40$~mm.  At even larger displacements the deviation rapidly grows, causing tune shifts and limiting the ring acceptance when the ion oscillations around the central orbit reach these very high amplitudes.

The overall focusing effect of a quadrupole is determined by longitudinal integrals over the field gradients.  Within the orbital equations \cite{rossbach_schmueser_1994} it is described by $\int_{(\text{Elem.})}k_{x(y)}(s)ds$, where the local focusing strengths $k_{x(y)}(s)$ are, for ions of charge $ze$ and the nominal ion energy $E_0$, given by $k_{x(y)}(s)=-zeg_{x(y)}(0,s,0)/2E_0$ (non-relativistic, harmonic-focusing approximation).  The integrals extend over the quadrupole element including its fringe fields.  For a hard-edged element of length $l_0$ with the ideal quadrupole gradients $\hat{g}_{x(y)}$, the focusing strengths inside the element would be given by $k_x=-k_y=k=zeU/E_0R_0^2$ and vanish beyond the edges; the resulting focusing integrals were $\pm\,kl_0$ for $x$ and $y$, respectively.  For the realistic fields, the focusing integrals can be expressed with the same $k$ as for the ideal case, but using $l_{\text{eff}}= \hat{g}^{-1}\int_{(\text{Elem.})}g_x(0,s,0)ds$ instead of $l_0$.  From the finite-element calculations using the TOSCA code, we obtain $l_{\text{eff}}=0.212$~m.  Hence, the finite boundaries lead to an effective quadrupole length 6\% larger than $l_0$.

When tuning the storage ring, the effective length is used to derive the voltages $U_i$ for the quadrupole families $i$ such that, according to a MAD8 ion-optics calculation, the focusing strengths $kl_{\text{eff}}$ yield a desired working point $Q_x, Q_y$.  The quadrupoles were designed for maximum voltages of $\pm10$~kV; at the CSR energy of 300~keV per unit charge and a typical working point, $|U_i| \approx 5$~kV.

\subsection{All-ring ion beam tracking calculations}
\label{tracking}

\begin{table}[t]
  \caption{Betatron functions ($\beta_x$, $\beta_y$) and dispersion ($D_x$) in the center of the CSR straight sections, together with betatron tunes $Q_x, Q_y$ of the CSR.  Results from matrix calculations (MAD8 \cite{MAD8}) are compared to those from all-ring tracking calculations using TOSCA \cite{Tosca} and G4beamline \cite{G4beamline}.  The tracking calculations also yield the approximate ring acceptances $A_x, A_y$, which are given for zero-emittance ion beams. }   \label{tbl:elements}
  \centering
\begin{tabular}{cd@{~~~~~~~~}ddl}
\hline
 \multicolumn{1}{c}{Parameter}  & \multicolumn{1}{c}{MAD8}   
    &  \multicolumn{1}{c}{TOSCA} & \multicolumn{1}{c}{G4beamline} &  \multicolumn{1}{c}{Unit} \\
\hline
$\beta_x$ & 12.44 & 12.1 & 12.41  &m    \\
$\beta_y$ & 1.47  & 1.3  & 1.4    &m     \\
$D_x$     & 2.06  & 2.1  &  2.1   &m      \\
$Q_x$     & 2.59  & 2.60 &  2.60     \\
$Q_y$     & 2.59  & 2.61 &  2.62     \\
$A_x$     &       & 120  &  120   & mm\,mrad\\
$A_y$     &       & 180  & 170    & mm\,mrad \\
\hline
\end{tabular}
\end{table}

\begin{figure}[b]
    \centering
    \includegraphics[width=70mm]{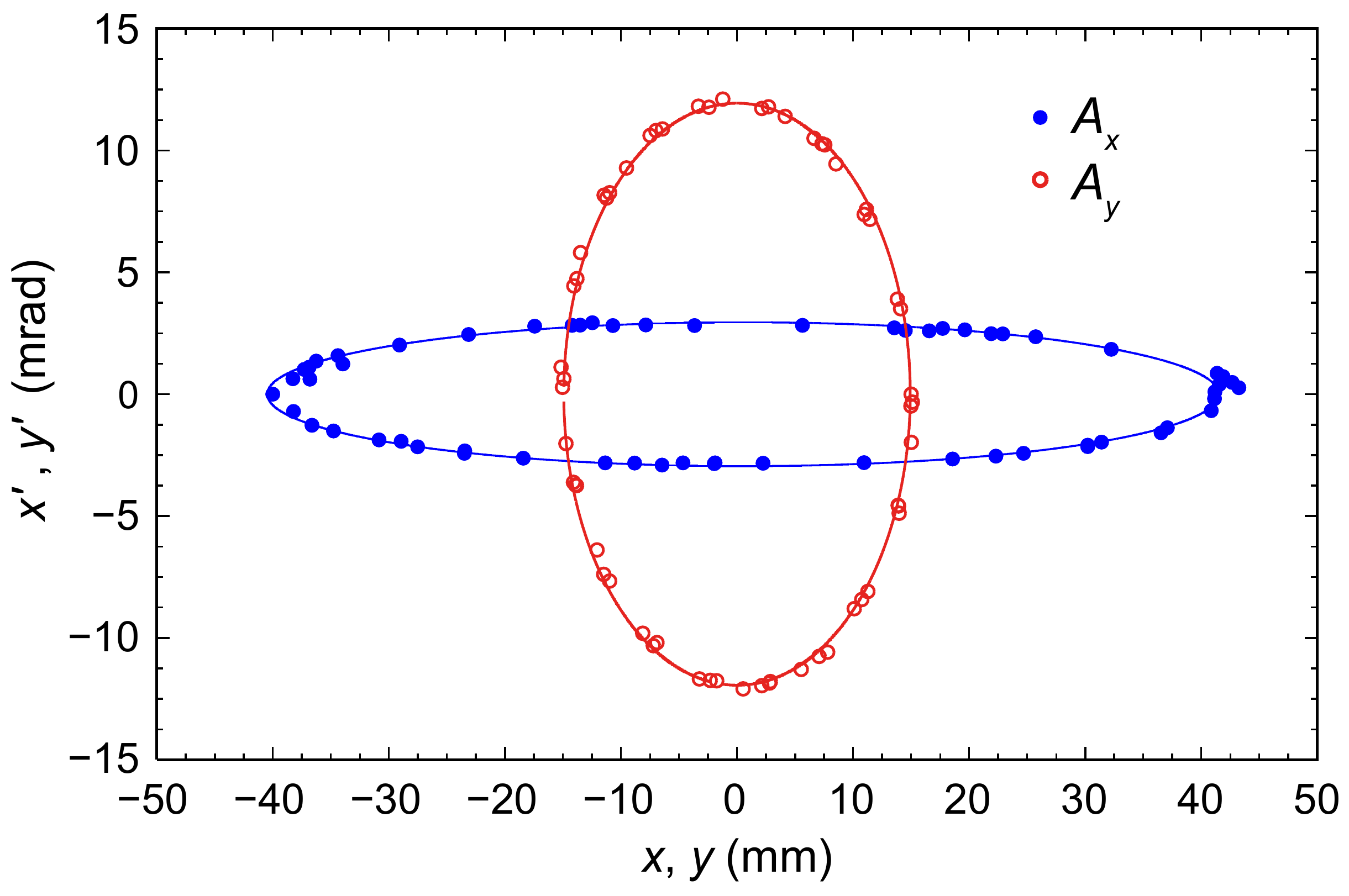}
    \caption{Tracking calculations performed using TOSCA for a particle started at the center of a straight section with $x=40$~mm, $y=15$~mm and $x'=y'=0$, showing the transverse positions and angles on subsequent turns at the same longitudinal position.  The starting momentum was the nominal storage-ring momentum $p_0$ ($\Delta p/p_0 = 0$).  The lines denote a fit of phase-space ellipses close to the acceptances $A_{x}$ and $A_{y}$ given in the text.  The maximum horizontal and vertical beam sizes at this position are found to be $\pm40$~mm and $\pm15$~mm, respectively.}
    \label{fig:acceptance}
\end{figure}

\begin{figure}
  \centering
    \includegraphics[width=70mm]{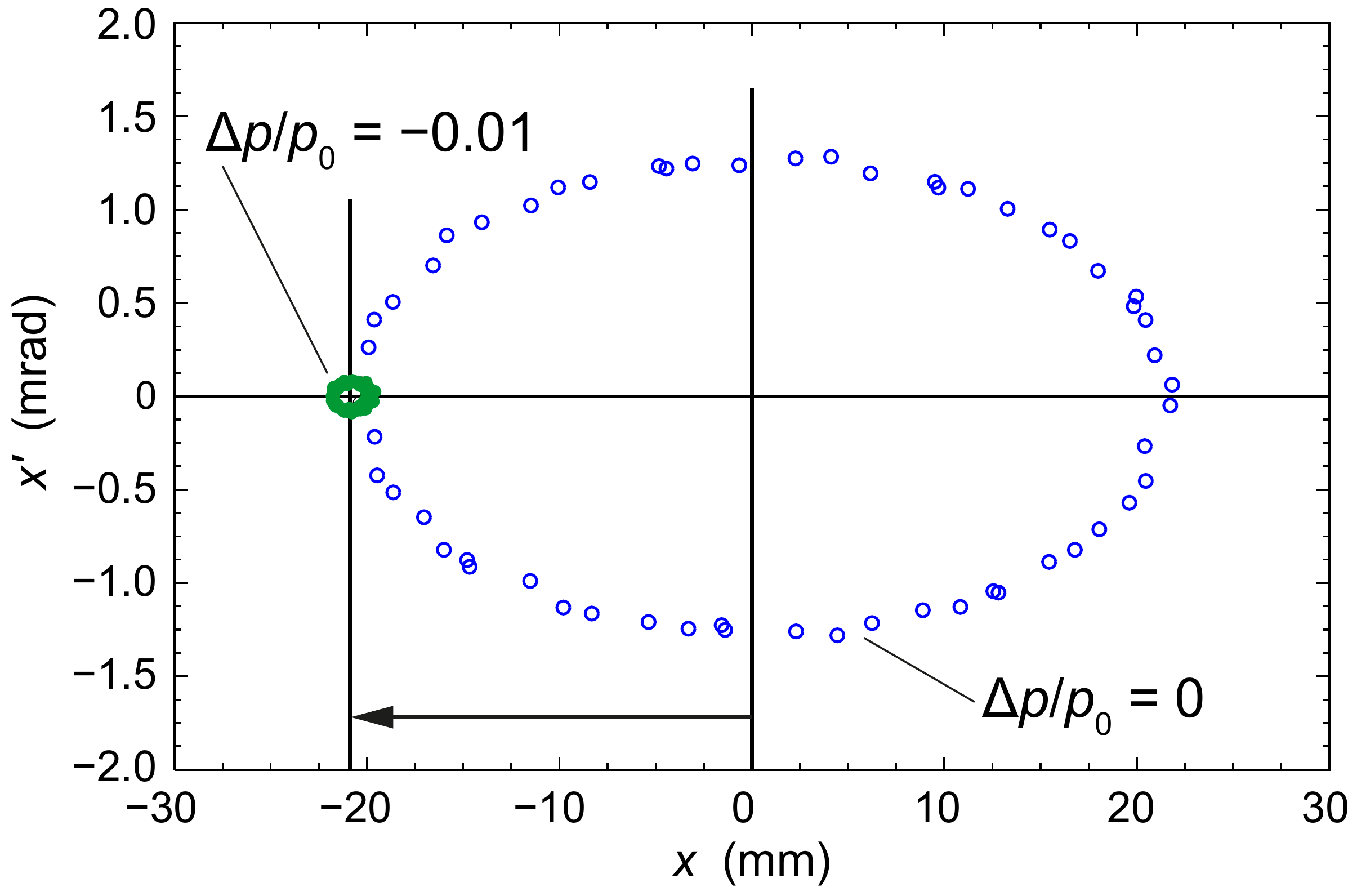}
    \caption{Tracking calculations by TOSCA started at the center of a straight section with $\Delta p/p_0 = 0$, $x=-21$~mm and
      $x'=y'=y=0$ (symbols on large ellipse) and $\Delta p/p_0 = -0.01$, $x=-20$~mm, $x'=y'=y=0$ (small ellipse near $x=-21$~mm),
      showing the horizontal particle positions and angles on subsequent turns at the same longitudinal position.  From the amount by
      which the center of the betatron oscillation shifts for the off-momentum particle (arrow), a value of $D_x =2.1$~m is found for
      the dispersion function at this orbit position.  For the chosen starting condition, the beam at $\Delta p/p_0 = -0.01$ has only a
      very small betatron oscillation amplitude.}
    \label{fig:dispersion}
\end{figure}

After calculating the field maps for each of the individual elements, a model of the entire ring was constructed in TOSCA to investigate the ion beam dynamics in the real fields of the CSR.  Once a complete solution was obtained, a simulated 300-keV proton was tracked through the CSR.  In Sec.~\ref{lattice} the CSR lattice parameters were found using the transfer matrix formalism of MAD8 \cite{MAD8}, where each element is represented by a user-defined matrix.  To compare these results with the TOSCA-simulated lattice parameters, we first used TOSCA to calculate, over several revolutions, the phase-space positions of a stored particle at the center position of a straight section.  The phase-space coordinates (transverse positions $x$, $y$ and angles $x'=dx/ds$, $y'=dy/ds$) were then fitted by an ellipse whose parameters represent the lattice parameters at this center position.  The lattice parameters obtained from MAD8 and from the detailed TOSCA calculations were found to be in reasonable agreement (see Table~\ref{tbl:elements}).

In further TOSCA calculations, the starting conditions were chosen to represent particles performing high-amplitude oscillations around the central orbit.  For oscillation amplitudes in the middle of a straight section of up to $\pm$40~mm in the horizontal and $\pm$15~mm in the vertical directions, regular elliptical phase-space maps of the horizontal and vertical betatron oscillations were found, as presented in Fig.~\ref{fig:acceptance}.  Inreasing the amplitudes further lead to irregular phase-space motion.  The critical amplitudes of $\hat{x}=40$~mm horizontally and $\hat{y}=15$~mm vertically can be rationalized considering the transverse extent of the good-field region in the quadrupoles ($\leq$\,40~mm, or 80\% of $R_0$, around the axis, see Fig.~\ref{fig:Qgradr}) in connection with the betatron functions in the center of a straight section (see Fig.~\ref{fig:beta}).  The smaller critical beam size in the vertical direction is due to the fact that the values of the betatron function in this direction at the quadrupole and in the middle of a straight section differ from each other by a factor of $\sim$\,4.5.  The horizontxal acceptance \cite{rossbach_schmueser_1994} is $A_{x}=\hat{x}^2/\beta_{x}=120$~mm\,mrad while, correspondigly, $A_y=180$~mm\,mrad.  It should be noted that $A_{x}$ and $A_{y}$ describe the acceptance limits (see Table~\ref{tbl:elements}) for beams with a fixed-amplitude betatron oscillation in a single coordinate only and that beam loss is expected to start already at significantly smaller beam emittances.

Tracking calculations with TOSCA were also performed with off-momentum particles.  By starting such particles at different relative momentum deviations $\Delta p/p_0$, the shift of the betatron-oscillation center with $\Delta p/p_0$ can be probed.  A value of $D_x=2.1$~m is deduced for the horizontal dispersion at the center of the straight section, in good agreement with the matrix calculations using MAD8 (see Table \ref{tbl:elements}).

Tracking calculations of the entire ring were also performed using the G4beamline program \cite{G4beamline}, where each element of the CSR was represented by the field table calculated with TOSCA.  The lattice parameters in the center of a straight section and the ring acceptances obtained by these calculations are in good agreement with those determined by the other methods.  The calculations by TOSCA and G4beamline show that the tracking of ions through the entire field geometry of a large electrostatic storage ring is feasible and can give valuable insight into the performance of the machine.

\subsection{Cryogenics and vacuum system}
\subsubsection{Mechanical layout}
\label{mechanical}

\begin{figure}
    \centering
    \includegraphics[width=86mm]{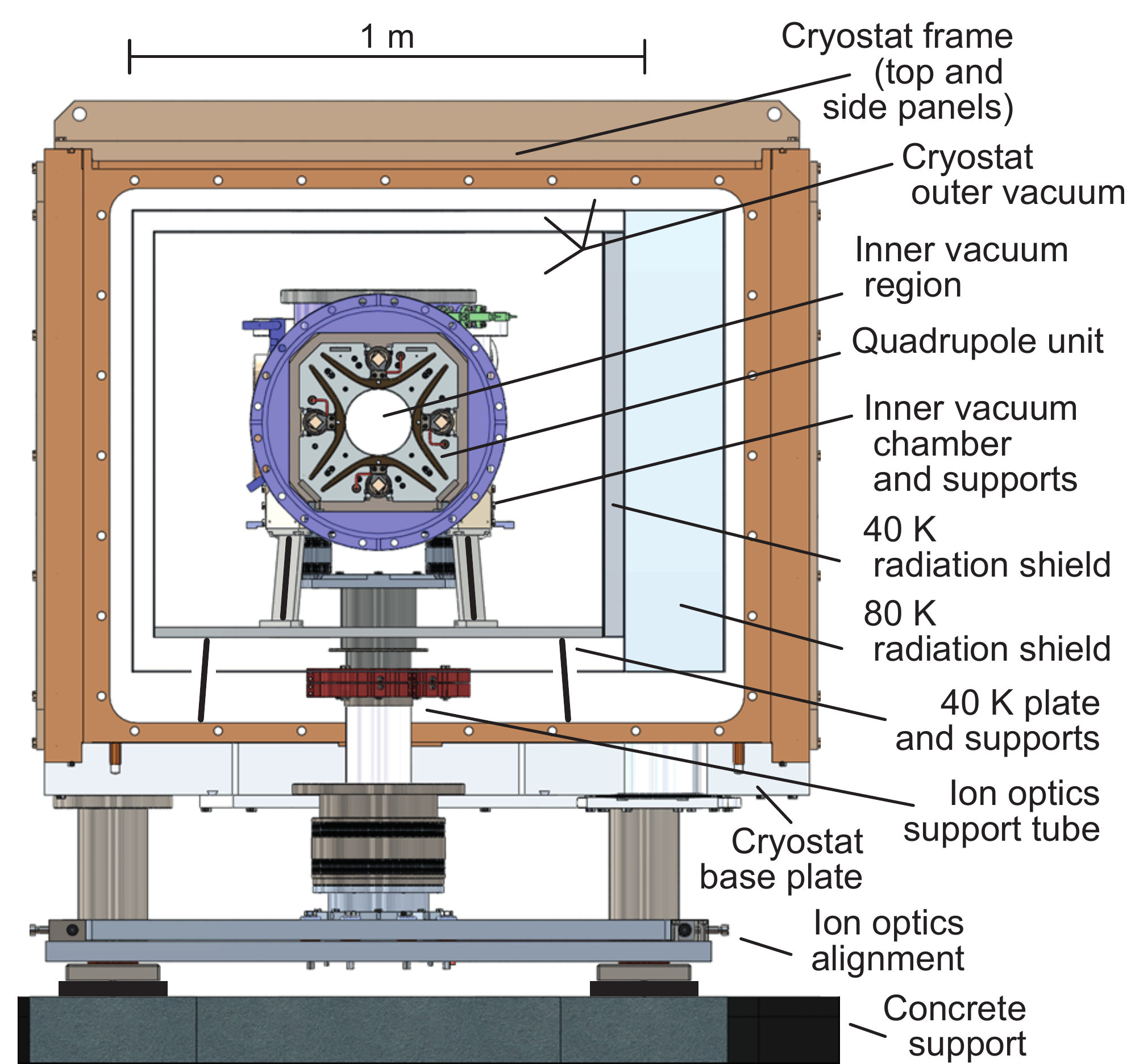}
    \caption{Sectional view of the CSR cryostat structure, looking against ion-beam direction into a corner section. The 40~K and 80~K shields open up to the right to accomodate the bend. The elements behind the quadrupole and the super-insulation are not shown.}
    \label{fig:CSRcrossec}
\end{figure}

The mechanical structure of the CSR consists of an ultrahigh vacuum (UHV) inner vacuum chamber (IVC) with 100 to 320~mm inner diameter (ID), kept at temperatures between 2 and 10~K.  The IVC is surrounded by inner and outer radiation shields at nominally 40 and 80~K, respectively, and an outer vacuum chamber (OVC) at, typically, $10^{-6}$~mbar.  The OVC has a square-toroidal shape that roughly follows the ion orbit and whose minor section is rectangular, offering a free internal cross section of at least $1\times1$~m$^2$ (see Fig.~\ref{fig:CSRcrossec}).  When the IVC is at room temperature, the OVC can be vented and opened while the IVC remains at UHV.  The IVC is then pumped by non-evaporable getter (NEG) modules placed close to the ion beam line and by ion getter pumps.  The ion getter pumps are located outside of the OVC and are operated at room temperature even when the IVC is cryogenically cooled.  With cryogenic cooling, the pressure in the IVC decreases.  Further details on the vacuum system and performance are given in Secs.~\ref{cryosystem}, \ref{vacsystem} and \ref{photodetach}.

The OVC comprises 16 large rectangular structures, each consisting of a stainless-steel (316L) frame and removable aluminum panels closing the frame from all sides.  The four corner sections and the four straight sections midway between them are supported by concrete blocks, with the straight sections separated from the corner sections on each side by hydroformed bellows with an ID of 1~m.

A vertical cross section of the cryostat and the elements mounted inside are shown in Fig.~\ref{fig:CSRcrossec}.  The room-temperature baseplates of the OVC, via thermally insulating supports, carry 15~mm-thick cold baseplates that form the bottom of the 40~K shield.  The cold base plates, in turn, carry most parts of IVC using similar thermally insulating supports.  The 144 to 165~mm high supports are made from 0.5~mm-thick, corrugated Ti sheet (Ti\,6Al\,4V, effective cross section per support 36.1~mm$^2$).  The IVCs, including their flanges, are produced from 316L stainless steel and sealed by Helicoflex \cite{Technetics} Cu-coated helical spring gaskets.  

The 40~K radiation shield, made from aluminum of type Al\,99.5, is mounted on the 40~K baseplate once the inner assembly is completed.  The 80 K shield, made from the same material, is attached below the 40~K baseplate and surrounds the remaining 40~K shield.  The 80~K shield is covered on the outside by 30--40 layers of super-insulation made from aluminized polyester film separated by polyester fiber spacers.

While the 40~K baseplates carry the IVC, all the beam focusing and bending elements of the CSR are mechanically decoupled (as much as possible) from the IVC by mounting them individually on external concrete support blocks.  For this purpose, a vertical tubular structure (stem) with thermally insulating sections extends from inside the IVC down to a room-temperature alignment structure on the support blocks.  Close to the IVC, the stems are made from thin-walled stainless-steel tubes in a meander-type mounting, while in their lower part they consist of a glass-fiber (G10) tube.  Between the vacuum zones, the stems pass through thin-walled bellows: small ones that are sealed to the IVC and larger ones sealed to the OVC.  This way, the electrodes remain fixed in the horizontal plane in spite of thermal contraction of the IVC.  The thermal contraction of all stems was found to lower the focusing plane of the ring by not more than 1~mm at cryogenic temperatures as compared to its position at room temperature.  Especially with respect to the ion optical elements, the mechanical structure of the CSR has been greatly modified from that of the cryogenic ion-beam trap, CTF \cite{lange_cryogenic_2010}, previously set up as a prototype for the CSR.

The electrodes were machined (cut and wire-eroded) from aluminum of type Al\,Mg4.5\,Mn0.7, electroplated by Ag and Au layers facing the beam, and mounted on alumina spacers for electrical insulation.  The electrical connections were made via alumina vacuum feedthroughs by 2.6--3.5~mm dia Cu wires ($>$\,99.9\% Cu or better) which are thermally connected to the IVC, using sapphire supports designed to hold the high operating voltages.  The outer electrical connections use stainless steel wires, thermally anchored to the 40~K shield by similar sapphire posts.  In electrical tests under room-temperature vacuum, the stable operation of this connection scheme was confirmed for high voltages up to those for ion energies of 300~keV per unit charge.

The deflecting electrodes of the CSR were aligned to a grid of retroreflective targets in the laboratory hall.  Other retroreflective targets were temporarily mounted to reference positions at the electrode assemblies and then, for alignment, observed by laser trackers.  The survey was first performed at room temperature.  In one corner section, a separate alignment procedure was performed through optical windows in the cryostat chambers.  This made it possible to repeat the survey with laser trackers also after the cooldown of the inner structures to $\sim$\,100~K.  The electrode positions at 100~K were found to be within $\pm$0.4~mm of their design values as well as from the actual positions at 300~K (with the small overall vertical shift due to thermal shrinking of all stems subtracted).  From beam dynamics calculations we found that, typically, an electrode deviation of 0.5~mm translates into a closed orbit shift of 1~mm.  Hence, effects of electrode misalignments are expected to be tolerable.  Moreover, this cold survey measurement showed that the concept of anchoring the cold electrodes to individual room-temperature supports worked successfully.

The quadrupole electrodes were aligned by a Taylor-Hobson telescope to bull's-eye targets outside the CSR.  Similarly to the deflectors, the quadrupole positions and alignment were confirmed both at cryogenic and at room temperature.

\subsubsection{Cryogenic system and performance}
\label{cryosystem}

The layout of the cryogenic system was governed by the goal of reaching extremely low pressures in the IVC.  The pressure limit relevant here is dominated by the vapor pressure \cite{honig_gases_1960} of H$_2$ in the coldest areas of the IVC.  Therefore, we introduced special locations along the circumference of the IVC with temperatures down to 2~K.  Surfaces serving this purpose were produced by vacuum brazing massive Cu blocks, connected to the coldest part of the cryogenic cooling circuit, into stainless-steel flanges.  From these blocks (Cu-HCP, 99.95\% Cu), finned surfaces \cite{von_hahn_electrostatic_2011} with 142~mm dia reach into the IVC, forming local cryocondensation units.  A total of 28 of these units are distributed over the CSR circumference at nearly-equal distances.

The closed-cycle liquid He cooling system of the CSR is supplied by a powerful refrigerator system described in more detail elsewhere \cite{von_hahn_electrostatic_2011,hahn_cryogenic_2006}.  Normal or superfluid He at a temperature of down to $\sim$\,1.8~K is fed into the finned Cu blocks and evaporated there into a return line for cold gaseous He.  This line is heat-exchanged with another one with He gas that enters the OVC at $\sim$\,5~K and is recirculated four times all around the ring to cool critical cryogenic equipment (at nominal 5~K level) and then, subsequently, the two radiation shields at 40~K and 80~K nominal temperatures.  The system \cite{hahn_cryogenic_2006} delivers a cooling power of 20 W to the 2~K level and 600~W to the radiation shields and experimental equipment.

The outside surfaces of the IVC are wrapped by a 0.25~mm-thick high-purity ($>$\,99.95\%) Cu sheet for reducing thermal gradients.  This layer is pressed to the stainless-steel wall of the IVC (nominal temperature $\lesssim$\,10~K) by collars at many places and connected to the Cu blocks of the pumping units, using highest purity (99.997\%) Cu strips.  The temperatures of the IVC and the radiation shields are monitored on the Cu layer by a large number of PT1000 sensors (becoming imprecise below $\sim$\,35~K) and a smaller number of RhFe sensors for the lowest temperature range.

\begin{figure}
    \centering
    \includegraphics[width=79mm]{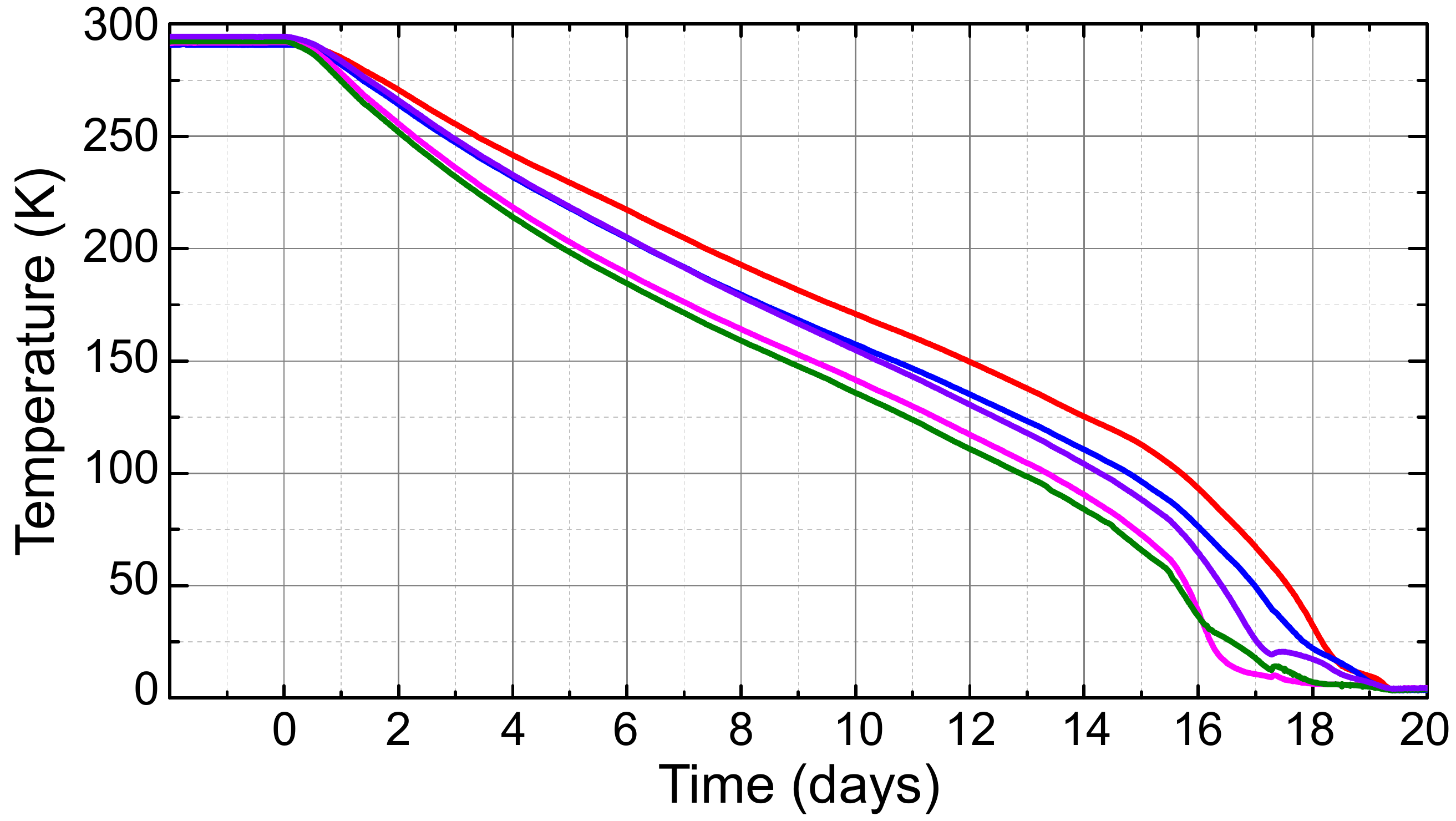}
    \caption{Temperatures recorded during the cooldown of the CSR at five RhFe probes (full lines) distributed over the inner vacuum chamber.}
    \label{fig:cooldown}
\end{figure}

In tests of a single corner, the cryocondensation units were brought to $(2.3\pm0.4)$~K and the temperatures on the IVC were found to be below 7.5~K.  The electrode temperatures were found to be $<$\,15~K in these tests.  As a result of these measurements, improvements were made in the thermal coupling of the electrodes to the IVC so that the temperature difference between them is now expected to be reduced to $\leq$\,5~K.

The first complete cryogenic cooldown of the CSR followed a bakeout of the IVC (see Sec.~\ref{vacsystem}).  The closed-cycle cryogenic system was regulated to realize a steady ramp-down of the temperatures with the deviations between the various PT1000 and RhFe sensors not exceeding 20~K from the mean.  Temperatures at some probes on the IVC during this regulated cool-down ramp are shown in Fig.~\ref{fig:cooldown}.  The minimum temperature was reached after $\sim$\,20 days.  It was obtained with the cryocondensation units at 4.2~K, as determined from the pressure in the return line of the evaporated He to the refrigeration system.  The mean temperature on the IVC was then determined to be $(5.5\pm1)$~K, where a large contribution to the uncertainty comes from the sensor calibration.  For the electrodes, which we expect to be at most 5~K warmer than the IVC, we hence estimate a temperature of $<$\,11.5~K.

\subsubsection{Vacuum system}
\label{vacsystem}

A goal followed in the design of the CSR inner vacuum system was to obtain good UHV conditions already at room temperature.  This ensures clean surfaces and low contributions of heavier gases in the residual gas during the cooldown, when the freeze-out on the cryogenic pumping surfaces occurs.  Moreover, these conditions allow ion storage with expected beam lifetimes in the range of several seconds and the operation of the merged-beams devices already at room temperature.  To achieve these UHV pressures, the IVC can be baked by electric heaters on the Cu wrapping at up to 250$^{\circ}$C with vacuum in the OVC.

The IVC is connected to five room-temperature pumping stations, installed below the corner sections, via 100~mm ID pumping ports, using bellows for thermal insulation.  In the cold state, the cross sectional areas of the pumping ports can be largely reduced by closing flaps connected to the 40~K temperature level.  Each room-temperature pumping station is equipped with a turbomolecular pump (H$_2$ pumping speed 220 l/s, isolated by gate valves after bakeout) and an ion getter pump \cite{starcell} of 65 l/s pumping speed (nominal value for N$_2$).  Inside the IVC there are 16 NEG modules with at least 700~l/s pumping speed each (for H$_2$, at room temperature) are used.  These modules are activated by heating to $\sim$\,450$^{\circ}$C in the final phase of the bakeout.  Moreover, specially designed charcoal-coated and bakeable cryopanels \cite{itep} are installed at eight positions for adsorbing heavier gases.  These cryopanels are connected to the refrigerating system at the 5~K level.  Finally, the 28 cryocondensation units directly connected to the liquid He cooling system take over cryocondensation of H$_2$ in the cold system and ensure extremely low residual-gas density in cryogenic operation.  This combination of vacuum pumps is essentially the same as in the CSR prototype \cite{lange_cryogenic_2010}, where it was shown to attain residual-gas densities of $2\times10^{3}$~cm$^{-3}$.

During the bakeout of the IVC, the temperatures were ramped up by 10~K/h below 100$^{\circ}$C and by 5~K/h above this.  The maximal temperature was 180$^{\circ}$C, sustained for 3 days and ramped down to 150$^{\circ}$C before the NEG pumps were activated.  After cooldown to room temperature, the pressure in the IVC, measured at the pumping stations below the corner sections with all turbomolecular pumps isolated, was $2\times10^{-10}$~mbar.  Before starting the cryogenic cool down, the OVC was opened to install heat-sensitive equipment such as cryogenic amplifiers of the diagnostic system.

The OVC is pumped down from atmosphere by two oil-free roots pumps.  These pumps are disconnected at $\sim$10$^{-2}$ mbar and the vacuum is then maintained by four turbomolecular pumps (1500~l/s each) and their respective oil-free backing pumps.  The pressures measured outside the super-insulation layers were $\sim$\,10$^{-5}$~mbar with the IVC at room temperature and 10$^{-7}$~mbar with the IVC at $<$\,100~K. 

\section{Experimental equipment}

The CSR is designed to house a wide range of experimental equipment.  For the first cryogenic operation of the facility, reported here, one of the straight sections was equipped with a laser-overlap section and with particle detectors which are described below.

\subsection{Basic considerations}

The laser setup implemented at the CSR in the first cryogenic experimental phase fulfilled two purposes.  Firstly, a laser system with a wide tuning range for dedicated photodissociation and photodetachment experiments was needed.  Secondly, several experiments with negative ions relied on a simple continuous laser system to create a measurable photodetachment signal that scaled with the number of stored ions.  A laser probe was necessary because of the extremely high vacuum anticipated for the CSR.  In most conventional, non-cryogenic storage rings and ion beam traps the residual-gas number densities are $>10^{6}$~cm$^{-3}$.  At such densities, ion collisions with the residual gas limit the lifetime of the stored beam.  Despite this limit, the residual-gas can also be a useful tool to monitor the beam decay and position.  For example, under such vacuum conditions, the stored particle beam will produce at a steady rate ionized residual-gas particles which can be directed onto microchannel plates that serve as beam-profile monitors \cite{hochadel_residual-gas_1994}.  It is also often possible to observe up- or down-charged ions that are created by charge exchange of the stored ions with the residual gas and leave the ion-beam orbit tangentially.  This process has been used routinely in magnetic \cite{schippers_electronion_2015} and room-temperature electrostatic \cite{andersen_physics_2004} heavy-ion storage rings to observe the beam lifetime and as a proxy signal that scales with the number of stored ions.

Inside the IVC during cryogenic operation, however, we expect extremely low number densities, several orders of magnitude lower than in traditional non-cryogenic storage devices.  Therefore, even considering that the charge-exchange cross sections increase at lower velocities, it was not clear whether we would be able to observe any neutral particles that are produced by residual-gas collisions.  As demonstrated in Sec.~\ref{photodetach} below, their detection required particle detectors with a low dark-count rate.

For anion beams, the most convenient method of monitoring the stored current was via photodetachment.  Most atoms in the periodic table can form stable negative ions.  However, the outer electron is bound by polarization forces only and, thus, the electron affinity (the binding energy of the extra electron) is typically much lower than the first ionization potential.  Therefore, a helium--neon (HeNe) laser at 633~nm is capable of neutralizing the majority of negative atomic ions in the periodic table, as well as many molecular anions.  This allowed very sensitive detection of stored negative ion currents by collecting the neutral detachment products in single-particle detectors downstream of the laser section (for details on detectors, see Sec.~\ref{detectors}).

\subsection{Ion--photon interaction zone}
\label{lasersection}
 
The laser setup was installed in the straight section of the CSR that will house the electron cooler at a later stage.  We used three different lasers in the first experimental phase.  The most versatile system is a pulsed optical parametric oscillator (OPO) laser (Ekspla NT342 with ultraviolet extension) with a wide tuning range from 2600~nm to 225~nm.  We used broadband-coated mirrors for most of the beam path to allow for rapid changes in wavelength without the need for complete re-alignment.  The pulse length of the OPO laser is 3~ns to 5~ns and the repetition rate 20~Hz.  The energy per pulse ranges from 30~mJ at 450~nm to 4~mJ at 2600~nm.  The linewidth is on the order of 5~cm$^{-1}$.  The other two systems are relatively simple, fixed-wavelength continuous (cw) lasers.  The first, a conventional HeNe laser with 5~mW power at a wavelength of 633~nm; the second, a home-built external-cavity-stabilized infrared diode laser with an output power of 20~mW at 1400~nm.  To switch between the different lasers we used flip-mirrors and a computer-controlled mechanical shutter.  For the data presented here we have employed the 633~nm cw laser.

The laser beams (typical diameter $\sim$\,8~mm) were introduced into the storage ring through sapphire viewports and small openings in the radiation shields.  The laser overlap with the stored ions was defined by two mirrors that were situated 224.3~cm apart inside the IVC on opposite sides of the ion-beam orbit.  In this arrangement, the laser beams intersected the ion beam at a grazing angle of 3.4$^{\circ}$.  In a second configuration, the laser beams could be coupled into the ring in a crossed-beam arrangement at a 90$^{\circ}$ angle.

\subsection{Particle detectors}
\label{detectors}

Neutral photodetached products were collected by one of the particle detectors placed downstream of the ion-photon interaction zone behind the minor deflector (see Fig.~\ref{fig:CSRlayout}).  Two detectors were used as follows.

The NICE (Neutral Imaging in Cold Environment) detector is based on a double-stack 120~mm diameter microchannel plate (MCP) with a phosphor screen anode.  For the counting experiments we used the digitized anode signal to find the time and amplitude.  The narrow pulse width of $\sim$\,6~ns full width at half-maximum (FWHM) made the NICE detector advantageous in the measurements with the pulsed OPO laser, where multiple events arising from a single laser shot reach the detector within a time spread of $\sim$\,1~$\mu$s.  The dark count rate of the large MCPs was greatly suppressed by correlating the ion arrival time with the time of the laser pulse.  This gating resulted in a background rate of $\lesssim$\,$10^{-4}$ per laser shot.  In future experiments, a fast CMOS camera will detect the transverse impact positions on the phosphor screen, thus providing full 3D-imaging multi-hit capability \cite{Urbain:RSI:2015}.

In the COMPACT detector \cite{Spruck:RSI:2015} (COld Movable PArticle CounTer) the detected particles hit an aluminum converter plate and the resulting secondary electrons are collected by a small MCP detector with a simple metal plate anode.  Since several secondary electrons are normally emitted per ion impact, the counting efficiency is not limited by the open area ratio of the MCP and can approach unity.  In cold operation the average dark count rate was substantially below 1~s$^{-1}$ even without any time-of-flight gating.  At average count rates up to $1\times10^{3}$~s$^{-1}$ no effects of detector saturation were observed in the experiments performed at cryogenic temperatures.  For these measurements, the COMPACT detector was operated without the additional local heating described in Ref.~\citenum{Spruck:RSI:2015}.  The high dynamic range of COMPACT was employed in the photodetachment measurements with the cw laser, presented below.  The detector is mounted on a translation stage by which it can be moved prependicular to the ion-beam axis in the horizontal plane.  This allows collection of not only neutral products (as for the present data), but also charged reaction products that leave the nominal ion orbit.  Both detectors operated successfully with the CSR at its lowest temperature and details of their operation will be reported separately.

\section{Ring performance and diagnostics}
 
The CSR was put into operation in March 2014, storing an $^{40}$Ar$^+$ ion beam at room temperature.  With a pressure in the $10^{-7}$~mbar range of the unbaked IVC, the storage lifetimes were limited to a few milliseconds.  Some beam diagnostic measurements were performed and the particle counters were tested.  In 2015, the IVC was baked to temperatures of $\sim$\,180$^{\circ}$C and a pressure of $2\times10^{-10}$~mbar at room temperature was reached, after which the IVC was cooled down to $\sim$\,6~K.  With the resulting extremely low pressures, several different ion beams were stored with storage lifetimes in the range of a few hundred to $\sim$\,2600~s.  The beam energies were in the range of 60--90~keV and the ions species included Ar$^+$, C$^-$, O$^-$, OH$^-$, CH$^+$, C$_2^-$, Co$_2^-$, Ag$_2^-$ and Co$_3^-$.  Exploiting the long storage lifetimes, extensive beam diagnostic measurements could be performed.  Moreover, investigations were carried out on spontaneous and photon-induced neutralization and fragmentation processes of the stored ions.  These studies used the particle detectors described in Sec.~\ref{detectors}.  The electron cooler was not yet available for this beamtime period.

\subsection{Ion sources and injection}
\label{injection}

The ions used in these experiments were produced in both positive and negative ion sources situated on an electrostatic platform designed to hold a maximum voltage of 300~kV.  Typical ion currents ranged from 1~nA to 1~$\mu$A. 

The injected beam was aligned to the storage-ring orbit by a beam viewer in front of the CSR and three further beam viewers that can be
moved into the CSR orbit.  The beam viewers consist of an aluminum plate on which secondary electrons are produced when hit by the ion beam.  Using a grid, the electrons are extracted and accelerated \cite{bergh_rex-isolde_2001} towards a 40~mm dia MCP and phosphor screen combination.  The image of the beam is recorded via a CCD camera.  For the CSR beam viewers, two cascaded rotary feedthroughs, one sealed to the OVC and the other to the IVC, are used to move the aluminum plate into the stored ion orbit.  In the operations performed so far, the basic setting of the ion-beam optical elements was found using the beam viewers at room temperature.  However, their operation at cryogenic temperature is foreseen.
 
In the injection beam line, a continuous (dc) ion current from the ion source is pulsed via a switched electrostatic deflector (chopper).  The first minor (6$^\circ$) deflector encountered by the ions, after entering into the CSR, serves as the active injection element of the ring.  Its voltages are switched off at the beginning of an injection cycle, emptying any ions from the storage ring.  Ions from the injector can then enter the CSR until the deflector voltages are once again raised to their nominal values within $\sim$50~ns, using solid-state switches, thus permitting ion-beam storage.  The ions that entered the CSR less than one revolution period before this switching time remain stored in the ring.  The leading edge of the stored pulse can be adjusted by setting the opening time of the chopper in the injection line, so that stored ion pulses of desired temporal length can be achieved.  The switching-on of the injection element produces the trailing edge of the circulating ion pulse.  This way the longitudinal current density distribution and, hence, time structure of the circulating ion beam is defined by the chopper and the injection element within the CSR.  The ``bunched'' time structure imprinted by the injection varies during storage and completely vanishes after a debunching time of typically 10$^4$ revolutions.

\subsection{Circulating ion current}
\label{current}

\begin{figure}
    \centering
    \includegraphics[width=82mm]{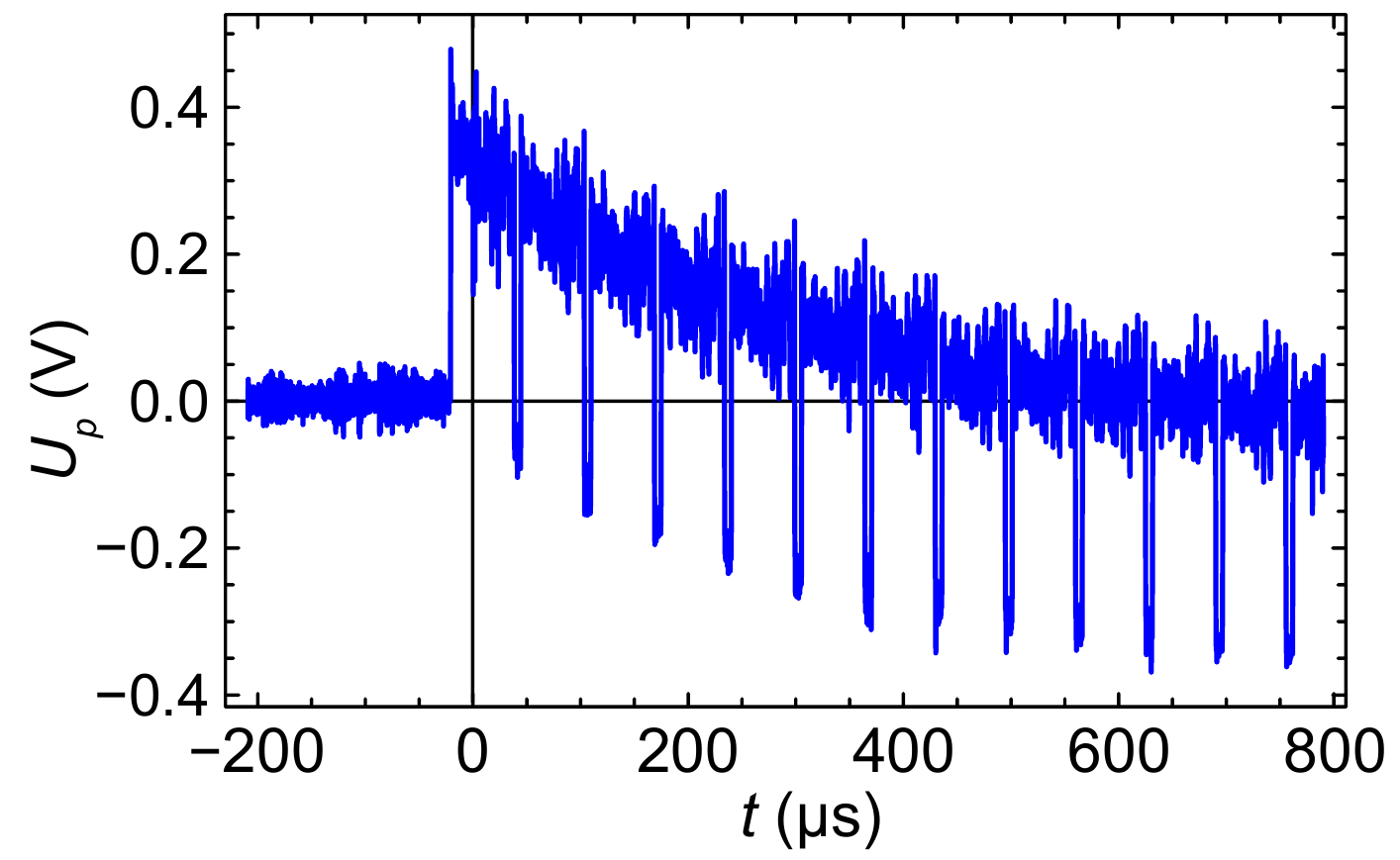}
    \caption{Time-domain signal at the current pickup observed for a 60~keV $^{40}$Ar$^{+}$ ion beam (revolution period $1/f_0=65.3$~$\mu$s) stored at $\sim$\,6~K, showing the amplified voltage $U_p(t)$.  At $t=0$ the injection element is switched on.  The trailing edge of the circulating ion pulse is observed in $U_p$ delayed by the time-of-flight from the injection element to the current pickup (slightly more than half a revolution period). Being recycled in the ring, the leading edge of ion pulse then arrives at the current pickup a second time.  For the case shown, the gap in the circulating current density is only $\sim$\,5~$\mu$s long.}
    \label{fig:Oszi}
\end{figure}
 
Two capacitive pickups \cite{boussard_diagnostics_1985} are used in the diagnostic section to detect the time structure of the circulating ion beam.  These pickups (100~mm ID tubes) differ in their length $l_p$.  The short pickup (current pickup PU-C, $l_p=30$~mm, see Fig.~\ref{fig:CSRlayout}) is employed to determine the charge associated with circulating ion pulses \cite{Grieser_diagnostics_2011}.  The long pickup (Schottky-noise pickup PU-S, $l_p=350$~mm) allows highest sensitivity for measurements of charge-density fluctuations in the circulating ion beam.  Even for a completely debunched ion beam, such fluctuations exist in the form of shot noise due to the single-particle nature of the stored beam current (Schottky noise \cite{boussard_diagnostics_1985}).  Such beam noise measurements are presented below (Sec.~\ref{noise}).

With the current pickup, in particular, the ion injection and the first few revolutions of the ions can be directly monitored.  The voltage $U_p(t)$ at this pickup as a function of time $t$ represents the linear charge density $\rho_l$ of the ion beam with a longitudinal resolution of $l_p/L_0\approx10^{-3}$ of the ring circumference ($L_0$).  The observed voltage, after amplification with a gain factor $g$, is $U_p=gl_p\rho_l/C$, where $C$ is the total capacitance of the pickup and the amplifier input. Aside from revealing the longitudinal distribution of the ions along the ring circumference, the pickup signal allows the number of stored ions (charge $ze$) to be determined as
\begin{equation}%
N_i=\frac{CL_0f_0}{zegl_p}\int U_p(t)dt.%
\end{equation}\label{eq:ni}%
Here, $f_0$ is the ion revolution frequency and the integral extends over one revolution period.  

The measured pickup signal for $^{40}$Ar$^{+}$ ions with a beam energy of 60~keV ($f_0=15.32$~kHz) is shown in Fig.~\ref{fig:Oszi}.  It reveals the pulse structure of a beam filling the circumference almost completely, as the chopper in the injection line is opened almost one revolution period before the voltages on the injection element are switched on.  The slow time dependence of $U_p(t)$ represents the high-pass filtering behavior of the capacitive pickup with a time constant of $RC$, where $R$ is the large amplifier input resistance.  The effective capacitance of the pickup electrode can be monitored in-situ by temporarily connecting a known inductance via a cryogenic relay and then measuring the resonance frequency of the combined circuit.  It was determined to be $C=(96.0\pm1.3)$~pF when operating the CSR at $\sim$\,6~K.  The signal was processed by a home-built cryogenic preamplifier stage operated at $\sim$\,6~K and a room-temperature main amplifier in series.  For the present experiments, the voltage amplification can be specified as $g=0.9^{+1.3}_{-0.1}\times10^3$.  The preamplifier was characterized both at the cryogenic operating temperature and at room temperature, which resulted in amplifications agreeing within $\sim$\,10\%.  The amplification $g$ was also measured directly during the cryogenic run for the complete chain and found to be about a factor of 2.5 lower than the product of the amplifications measured for the two amplifiers separately, when the input and output impedances were carefully simulated.  Since this deviation was not reproducible in later tests at room temperature, the upper error bar of $g$ includes the product of the two single amplifications.  The integral over a single pulse in Fig.~\ref{fig:Oszi}, ($2.0\pm0.1)\times10^{-5}$~Vs, then yields an ion number of $N_i=2.4^{+0.3}_{-1.3}\times10^8$ ions.  The noise integrated over a single revolution at the given $f_0$ corresponds to roughly $10^6$ ions ($z=1$), which characterizes the sensitivity of the pickup PU-C in these real-time measurements.  A calibration accuracy of $\sim$\,3\% is expected after suitable improvements.

\subsection{Ion-beam Schottky noise}
\label{noise}

The time structure imprinted by the injection process onto the stored ions vanished after $\sim$\,1~s of storage time.  Once the inner vacuum system reached its final cryogenic temperature, a signal from the Schottky noise of the stored ion beam could be observed up to very long storage times of 1000~s and more.  The signal of the Schottky-noise pickup (PU-S in Fig.~\ref{fig:CSRlayout}) is amplified by a low-noise, high-impedance preamplifier identical to the one used for the pickup PU-C, while a different room-temperature amplification stage is used.  The output signal is fed into a spectrum analyzer, where the spectral power distribution $\hat{p}(f)$ is observed at a harmonic $h$ of the ion revolution frequency $f_0$.  From calibration measurements, allowing for a similar uncertainty of the electronic readout as with PU-C, the total voltage amplification for frequencies of the order of 200~kHz can be specified as $g=8.2^{+0.7}_{-5.3}\times10^{3}$.  

\begin{figure}
    \centering
    \includegraphics[width=82mm]{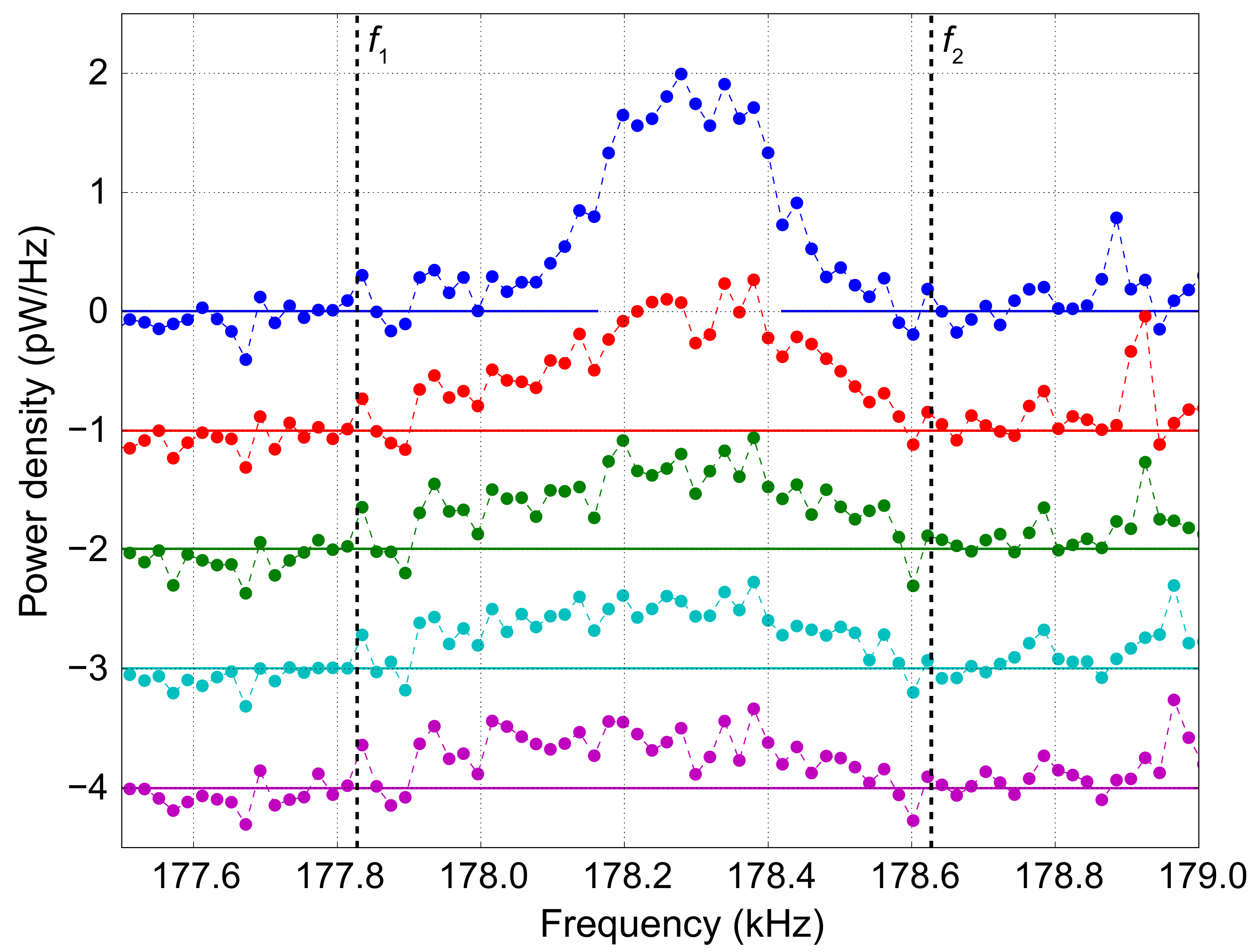}
    \caption{Measured Schottky-noise spectra of a Co$_{2}^-$ ion beam ($E_0=60.31$~keV, harmonic $h=20$) measured with averaging windows of 120~s centered at (from top to bottom) 80, 320, 560, 800 and 1040~s from injection, respectively.  For better visibility, the curves are shifted by $-$1 pW/Hz for each step of 240~s in storage time.  The amplification chain discussed in the text was used and a background spectrum taken with ion beam off was subtracted. Small peaks from frequency-variable ambient noise remain at the highest frequencies.  Limits $f_{1}$ and $f_{2}$ are indicated beyond which no ion beam related spectral signal is visible at any time.}
    \label{fig:Schottky}
\end{figure}

Power spectra of the amplified Schottky-noise signal for 60~keV Co$_2^-$ ions (mass 118~u, $f_0=8.915$~kHz) near the harmonic $h=20$ are shown in Fig.~\ref{fig:Schottky} for various storage time intervals.  From the Schottky-noise spectrum \cite{boussard_diagnostics_1985}, the spectral density of the average squared voltage at the pickup ($C=(452\pm15)$~pF) for a single ion (charge $e$, frequency $f_i$) in the Schottky band at $hf_i$ can be calculated as
\begin{equation}%
\langle\hat{u}^2\rangle = \left(\frac{e}{\pi hC}\right)^2 \left[1-\cos\left(\frac{2\pi hl_p}{L_0}\right)\right]\,\delta(f-hf_i),
\end{equation}\label{eq:spectraldensity}%
yielding the value of $2.2\times10^{-23}$~V$^2\,\delta(f-hf_i)$.  The amplified time-averaged noise power integrated over each harmonic (band $\Delta f$) per particle amounts to $\int_{(\Delta f)}\hat{p}(f) df = \int_{(\Delta f)} g^2\langle\hat{u}^2\rangle/R_m = 3.0^{+0.5}_{-2.6}\times10^{-17}$~W (with $R_m=50$~$\Omega$ being the spectrum analyzer input impedance).  The integrated power at 80~s (Fig.~\ref{fig:Schottky}) is $(5.5\pm0.5)\times10^{-10}$~W, corresponding to $0.18^{+1.2}_{-0.02}\times10^{8}$ Co$_{2}^-$ ions.  With the current pickup (see Sec.~\ref{current}) the initial number of ions was measured to be $N_i=0.8^{+0.1}_{-0.4}\times 10^8$.  In spite of the large uncertainties, the comparison confirms that the observed spectral power has the order of magnitude expected for a Schottky-noise signal.

\begin{figure}
    \centering
    \includegraphics[width=82mm]{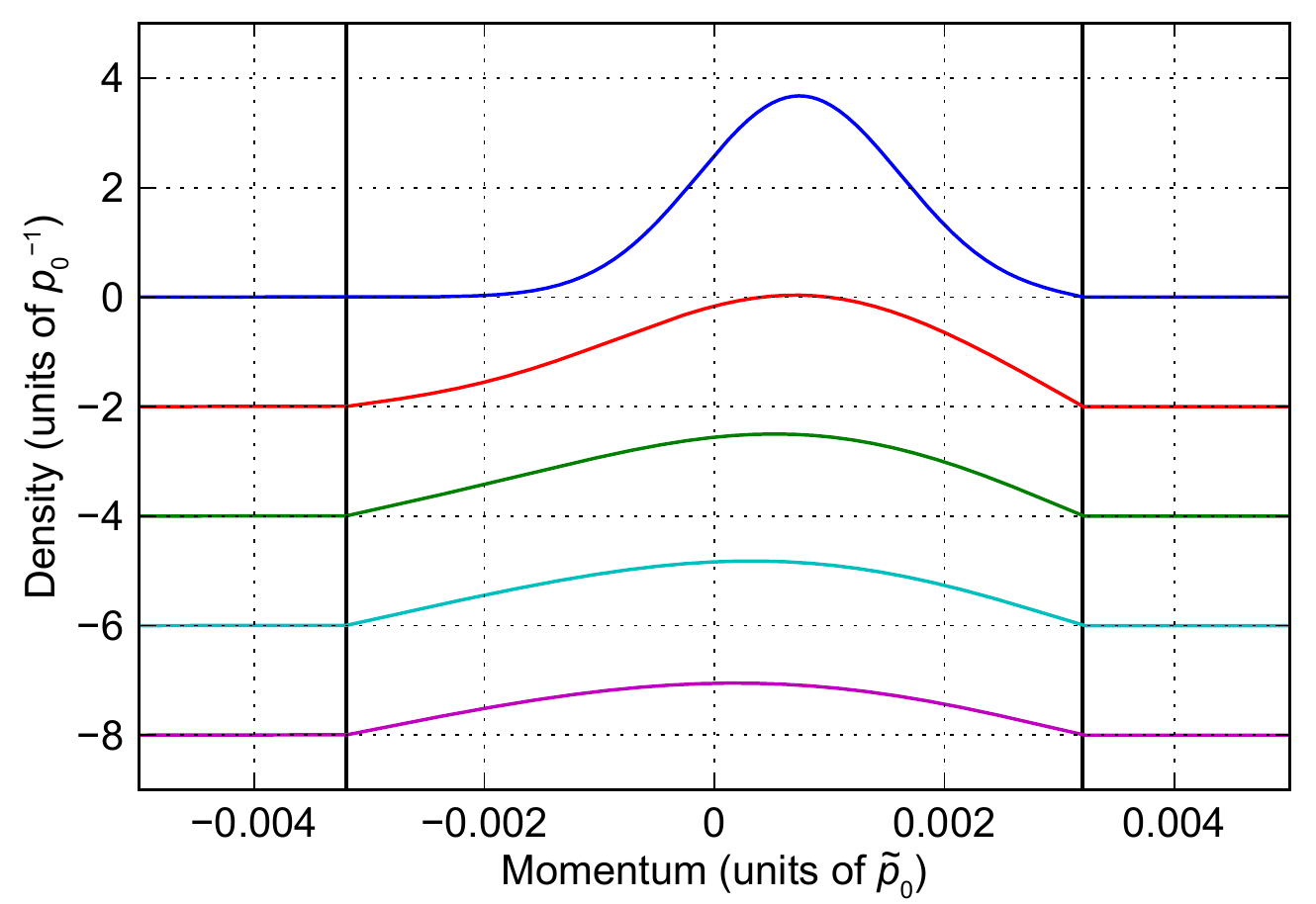}
    \caption{Momentum distributions (relative deviation from the reference momentum $\tilde{p}_0$) calculated with a linear diffusion model as discussed in the text to simulate the Schottky-noise spectra of Fig.~\ref{fig:Schottky}.  The five curves correspond to the average storage times of these spectra and for visibility are shifted by $-2\,\tilde{p}_0^{~-1}$ per step in storage time.}
    \label{fig:diffusion}
\end{figure}

The power spectrum $\hat{p}(f)$ represents the distribution of revolution frequencies in the stored beam and, hence, the distribution $w(p)$ of the longitudinal momenta of the stored ions.  For small deviations from a reference frequency $\tilde{f}_0$ or the corresponding reference momentum $\tilde{p}_0$, both distributions are related via 
\begin{equation}%
(f-\tilde{f}_0)/\tilde{f}_0 = \eta_p (p-\tilde{p}_0)/\tilde{p}_0.  
\end{equation}\label{eq:dfdp}%
The phase-slip factor $\eta_p$ was determined as $0.68\pm0.01$ by measurements and by the lattice calculations.  Figure~\ref{fig:Schottky} thus indicates that the momentum distribution of the ions significantly broadens during the long storage time.  Moreover, with time the integral power of the Schottky band shows a near-exponential decrease.  A time constant of $(1115\pm 33)$~s can be fitted to the observed Schottky-power decrease, which is interpreted as a decrease in the number of stored ions.  

Applying the Fokker--Planck description \cite{moehl_1980} of the longitudinal momentum distribution $w(p,t)$ we have simulated both the broadening of this distribution and the loss of ions from the beam by solving the linear diffusion equation \cite{selvadurai00}
\begin{equation}%
\partial w/\partial t = D (\partial^2 w/\partial p^2).
\end{equation}\label{eq:diff}%
We place the reference momentum $\tilde{p}_0$ where it corresponds to the average $\tilde{f}_0=(f_1+f_2)/2$ of the limits indicated in the Schottky-noise spectrum of Fig.~\ref{fig:Schottky} and consider it as the nominal (soll) momentum given by the voltage settings of the CSR.  Symmetrically around $\tilde{p}_0$ we define effective momentum acceptance limits $p_{1,2}=\tilde{p}_0\mp\Delta p/2$ with $\Delta p = \tilde{p}_0(f_2-f_1)/\eta_p\tilde{f}_0=0.0064\,\tilde{p}_0$ and use the boundary condition $w(p_1,t)=w(p_2,t)\equiv0$ in the diffusion model.  The starting conditions for the momentum distribution and the diffusion constant are chosen to simulate the shape of the measured Schottky-noise spectrum by the modeled momentum distribution.  As shown in Fig.~\ref{fig:diffusion}, reasonable agreement can be obtained by choosing an initial Gaussian distribution of width $\sigma=0.45\times10^{-3}\,\tilde{p}_0$ with an initial displacement of $+0.74\times10^{-3}\,\tilde{p}_0$, corresponding to a slight deviation from the soll momentum on injection, and a diffusion constant of $D=3.58\times10^{-9}\,\tilde{p}_0^2$~s$^{-1}$.  Through the boundary condition, the solution of the diffusion equation for these parameters also implies an exponential decrease of the particle number with a time constant \cite{selvadurai00} of $(\Delta p)^2/\pi^2D=1160$~s, close to the time constant observed.

The simple model agrees quite well with the observations.  In particular, it can explain the slight shift of the average ion momentum with storage time revealed by the Schottky-noise spectra and the asymmetry of the distributions, especially pronounced in those centered at 320~s and 560~s.  At the low residual-gas density reached, the energy loss in collisions of the circulating ions with background-gas molecules cannot be responsible for the observed shift; using the results of Sec.~\ref{photodetach}, a relative momentum change of only $<$10$^{-6}$, even after 1000~s, can be estimated for this effect.  In the diffusion model the shift of the average momentum is caused by the effective acceptance limits when the ions are injected slightly off the nominal momentum $\tilde{p}_0$.  On the other hand, the model neglects important influences from the transverse beam size and the transverse momentum distribution.  Thus, it suggests an effective momentum acceptance somewhat smaller than the single-ion acceptance limits obtained from the lattice calculations.  Nevertheless, it appears plausible that the effective momentum acceptance would be critically reduced by the large transverse beam size, found to be at least 20~mm FWHM in position scans over photodetachment products of the stored beam with the movable COMPACT detector.

The root-mean-square (rms) change of the particle energy per revolution consistent with the diffusion constant $D$ would be $\langle (\delta E)^2 \rangle^{1/2} = 2E_0(2D/\tilde{p}_0^2\tilde{f}_0)^{1/2}=107$~meV. Assuming independent effects, added in quadrature, of the 32 main elements (deflector and quadrupole electrodes) of the CSR, this would correspond to an rms energy change of 19~meV per element.  The origins of the frequency broadening of the observed Schottky signals will be further studied in future beamtimes.

\subsection{Ion-beam bunching}

A sensitive method for detecting weak stored ion beams is to re-bunch the stored ion beam using a radiofrequency (rf) system.  The rf system (see Fig.~\ref{fig:CSRlayout}) consists of a 340~mm long drift tube that is powered by an alternating voltage with a typical amplitude of 1--20~V at a higher harmonic ($h \sim 10$) of the revolution frequency $f_0$.  In the present experiments, the rf amplitude was zero when ions were injected into the ring and then ramped up to its final value over $\sim$\,20~ms for 60~keV $^{40}$Ar$^+$ ($f_0=15.32$~kHz), which corresponds to $\sim$\,300 revolutions. More details of the rf system are given elsewhere \cite{Grieser_diagnostics_2014}.  The rf bunching scheme is used in the measurements described in Sec.~\ref{positionmeas}.

\subsection{Ion-beam position and dispersion measurements}
\label{positionmeas}

The closed orbit of a bunched ion beam in the CSR can be measured with six beam-position monitors (BPM) distributed along the ring (see Fig.~\ref{fig:CSRlayout}).  A single BPM module consists of two 100~mm ID, 60~mm long cylindrical capacitive pickups, each split along their diagonal into a pair of separate electrodes \cite{Laux_2009}.  The normal of the splitting plane lies either in the horizontal or the vertical plane, yielding electrode pairs that are then called horizontally or vertically separated, respectively.  The electrodes are surrounded by a shielding tube that is electrically isolated from the IVC, but connected with it externally at a well-defined grounding position.  For the measurements described here, one electrode of a pair is connected to a single amplification chain, using a cryogenic relay, while grounding the other.  The measurement is then repeated with the connections interchanged and the center-of-charge of the ion beam in the BPM is derived from the difference of the signal amplitudes in the two measurements, normalized to their sum \cite{koziol_1994}.

Beam-position measurements were made to measure the closed orbit of the stored ion beam as well as to determine the dispersion of the storage ring at the pickup positions.  In these measurements, an $^{40}$Ar$^+$ ion beam was injected into the CSR and the closed orbit was then changed by varying all electrostatic potentials by the same relative amount $\Delta U/U$.  This leads to a horizontal shift $\Delta x$ of the center-of-charge position of the stored beam, which is related to the dispersion $D_x$ at the pickup position by $D_x=-2\Delta x/(\Delta U/U)$.  Since all BPMs are installed at equivalent positions within the ion-optical lattice of the storage ring, the dispersion at the six pickup positions should be identical.  The measurements yielded an average dispersion of $\overline{D}_x=2.20$~m, while the $D_x$ values measured at the individual BPMs varied in the range of 2.08~m to 2.28~m.  Calculations using the G4beamline code \cite{G4beamline}, where ions were tracked through the realistically modeled fields of the storage ring, yield for the used working point a dispersion of $D_x=2.14$~m, in reasonable agreement with the measured value.

\subsection{Betatron-frequency measurements}
\label{meas_betatron}

The induced current measured on a single electrode of one of a BPM's split cylinders corresponds to either the horizontal or the vertical motion of the ion-beam center-of-charge.  An oscillation of the center-of-charge position can be excited by injecting the ion beam with either a horizontal or a vertical offset relative to the central orbit.  When the frequency spectrum of such transverse pickup signal is analyzed for an ion beam still having the pulsed structure of the injection (see Sec.~\ref{injection}), sidebands at $f_{h\pm}^{x(y)}=f_0(h \pm q_{x(y)} )$ appear in addition to the main bands at $hf_0$.  Here, $x$ or $y$ apply to an observation pickup with electrodes separated in horizontal or vertical direction, respectively, $h$ is integer, and $q_{x(y)}$ are the non-integer parts of the horizontal and vertical tunes ($q_{x(y)}=Q_{x(y)}- \lfloor Q_{x(y)} \rfloor$).

\begin{figure}
    \centering
    \includegraphics[width=80mm]{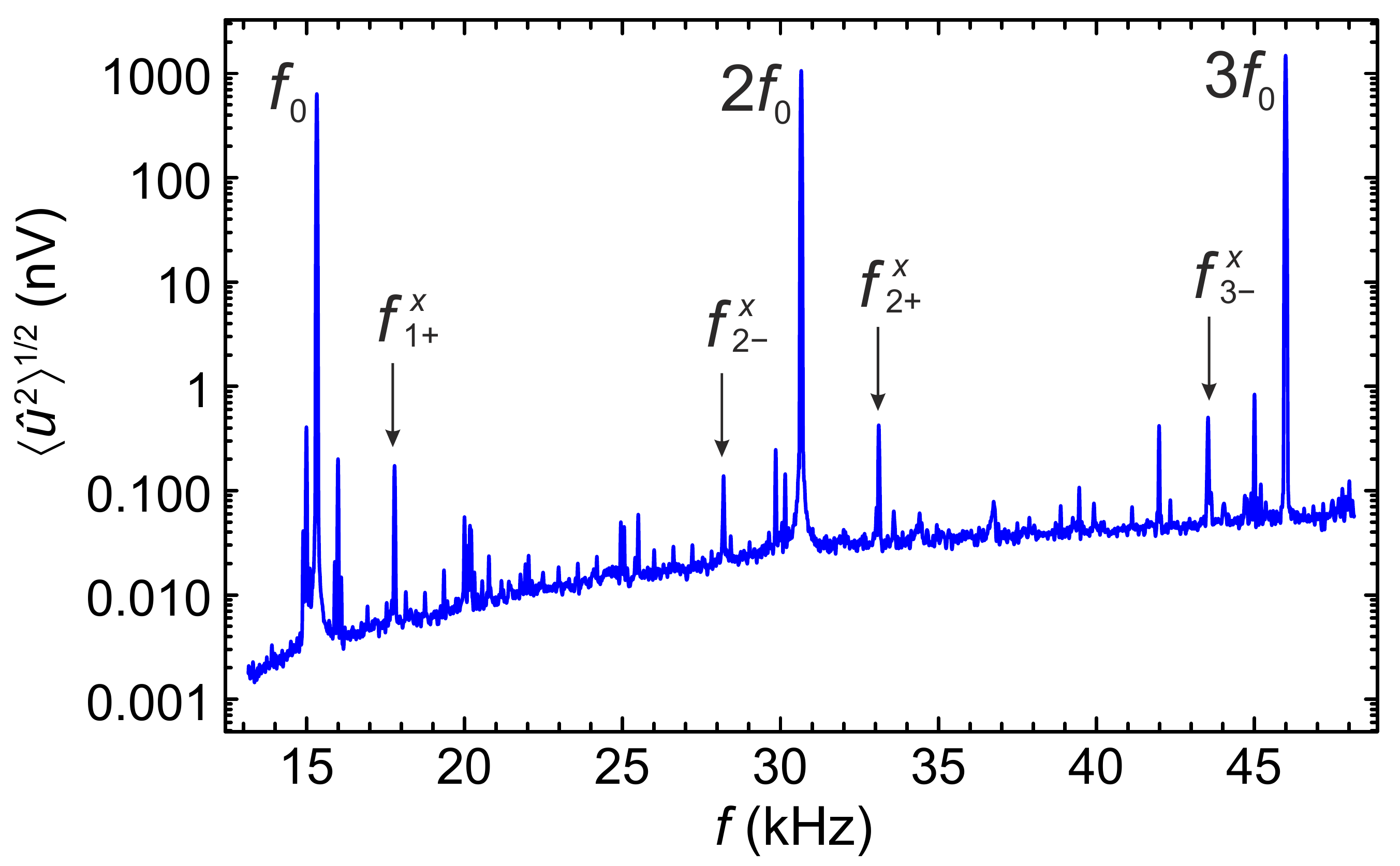}
    \caption{Betatron sidebands observed on a single electrode of a horizontally separated pair of a BPM for a 60~keV $^{40}$Ar$^+$ beam ($f_0=15.32$~kHz) injected with slight offset to the closed orbit.  The spectral amplitude $\langle\hat{u}^2\rangle^{1/2}$ is shown for 20~Hz bandwidth.  Harmonics of the revolution frequency $f_0$ are marked as $hf_0$ and the betatron sidebands as $f_{h\pm}^x$.}
    \label{hplate}
\end{figure}

Betatron sidebands of this type were observed in the pickup spectrum during the first few seconds after beam injection into the CSR.  For a 60~keV $^{40}$Ar$^+$ ion beam with $f_0=15.32$~kHz, a spectrum was measured for a horizontal injection misalignment, the signal of which, measured with one of the horizontally separated electrodes of a BPM, is displayed in Fig.~\ref{hplate}.  Focusing settings slightly off the nominal working point of the CSR were chosen for this measurement.  The highest three peaks in the spectra belong to integer multiples of the revolution frequency, reflecting the longitudinal density modulations of the freshly injected beam, while the small peaks marked with $f_{h\pm}^x$ are the betatron sidebands.  These peaks have similar amplitude to several noise peaks occurring in the spectrum.  However, they can be identified by their multiple occurrence at positions symmetrically correlated to the $hf_0$ peaks.  The data of Fig.~\ref{hplate} yield a horizontal tune of $Q_x=2.840\pm0.001$, where the integer part of $Q_x$ is set to its value determined from simulations, and a similar measurement yields $Q_y=2.397\pm0.001$ for the vertical tune.  At the given ring geometry, the MAD8 lattice calculations accurately predict the relation of the tunes $Q_{x(y)}$ to the focusing strengths of the quadrupoles.  By relating the focusing strengths that produce the measured $Q_{x(y)}$ values to the actually applied quadrupole voltages $U_i$ in each family $i$ (see Sec.~\ref{focquads}) effective lengths $l_{{\rm eff},i}$ can be experimentally determined.  They are found to be $l_{{\rm eff},1} = (0.207\pm0.004)$~m and $l_{{\rm eff},2} = (0.209\pm0.004)$~m, respectively.  As expected, these values lie very close to each other and to the finite-element calculations of Sec.~\ref{focquads}.

\subsection{Measurement of the horizontal and vertical betatron functions}

A well-known expression \cite{courant_theory_1958} shows that small changes $\Delta k_{x(y)}(s)$ of the focusing strength at longitudinal positions $s$ along a storage ring lead to changes of the betatron tune by
\begin{equation}%
\Delta Q_{x(y)} = \frac{1}{4\pi} \int\beta_{x(y)}(s)\Delta k_{x(y)}(s)ds
\end{equation}\label{eq:dq}%
where the integral runs over the ring circumference.  (With the definition of Sec.~\ref{focquads}, $k_x=k$ and $k_y=-k$.)  The effect, due to the eight members of each quadrupole family $i$, when changing the quadrupole voltage by $\Delta U_i$ can hence be obtained as
\begin{equation}%
\Delta Q_{x(y),i}=\pm 2zel_{{\rm eff},i}\overline{\beta}_{x(y),i}\Delta U_i/\pi E_0R_0^2
\end{equation}\label{eq:dqfin}%
where the plus (minus) sign refers to $x$ ($y$) and $\overline{\beta}_{x(y),i}$ is the average of the betatron functions at the individual quadrupoles of family $i$.  The respective individual values within a family should all be equal considering the ring symmetry.

\begin{table}
  \caption{Betatron function values $\overline{\beta}_{x(y),1}$ and  $\overline{\beta}_{x(y),2}$ at the quadrupoles of families 1 and 2, respectively, determined from the effect of these elements on the CSR betatron tune.  The measured values are compared to the values calculated by the transfer-matrix model (MAD8) at the quadrupole positions.}  \label{tbl:betam}
  \centering
\begin{tabular}{cr@{$\,\pm\,$}lcl}
\hline
 \multicolumn{1}{c}{Parameter} & \multicolumn{2}{c}{Measured} & \multicolumn{1}{c}{Calculated} & \multicolumn{1}{c}{Unit} \\
\hline
$\overline{\beta}_{x,1}$ & 9.55 & 0.23 & 8.69 & m \\
$\overline{\beta}_{y,1}$ & 3.49 & 0.07 & 3.41 & m \\
$\overline{\beta}_{x,2}$ & 4.20 & 0.03 & 4.11 & m \\
$\overline{\beta}_{y,2}$ & 6.64 & 0.13 & 6.37 & m \\
\hline
\end{tabular}
\end{table}

The betatron functions at the quadrupole positions can, hence, be determined from betatron tune measurements varying the quadrupole voltages $U_i$.  The results for the working point of $Q_x=2.840$ and $Q_y=2.397$ (see Sec.~\ref{meas_betatron}) are shown in Table~\ref{tbl:betam} and compared to the calculated betatron functions for this working point, which slightly differs from that in Fig.~\ref{fig:beta}.  The measured and calculated betatron functions deviate by at most $\sim$\,10\%, suggesting that the matrix calculations represent the real ion optics with good accuracy.

\subsection{Lifetime measurements and negative ion detachment}
\label{photodetach}

During the cool-down period of the CSR, the lifetime of a stored 60~keV $^{40}$Ar$^+$ ion beam was frequently measured.  In this phase of the measurements, ion collisions with residual-gas molecules in the inner vacuum system lead to neutral Ar atoms via electron capture.  These atoms were detected by the COMPACT detector (see Sec.~\ref{detectors}) as a function of storage time.  Starting with a beam lifetime of $\sim$\,5~s at room temperature, the continuous improvement of the vacuum, by cryopumping, lead to longer and longer storage times.  In fact, reaching cryogenic temperatures of $\sim$\,6~K, the signal of fast atoms on the detector from the ions' collisions with residual gas dropped below the dark count rate of the detector, indicating a tremendously improved background pressure.

\begin{figure}
    \centering
    \includegraphics[width=82mm]{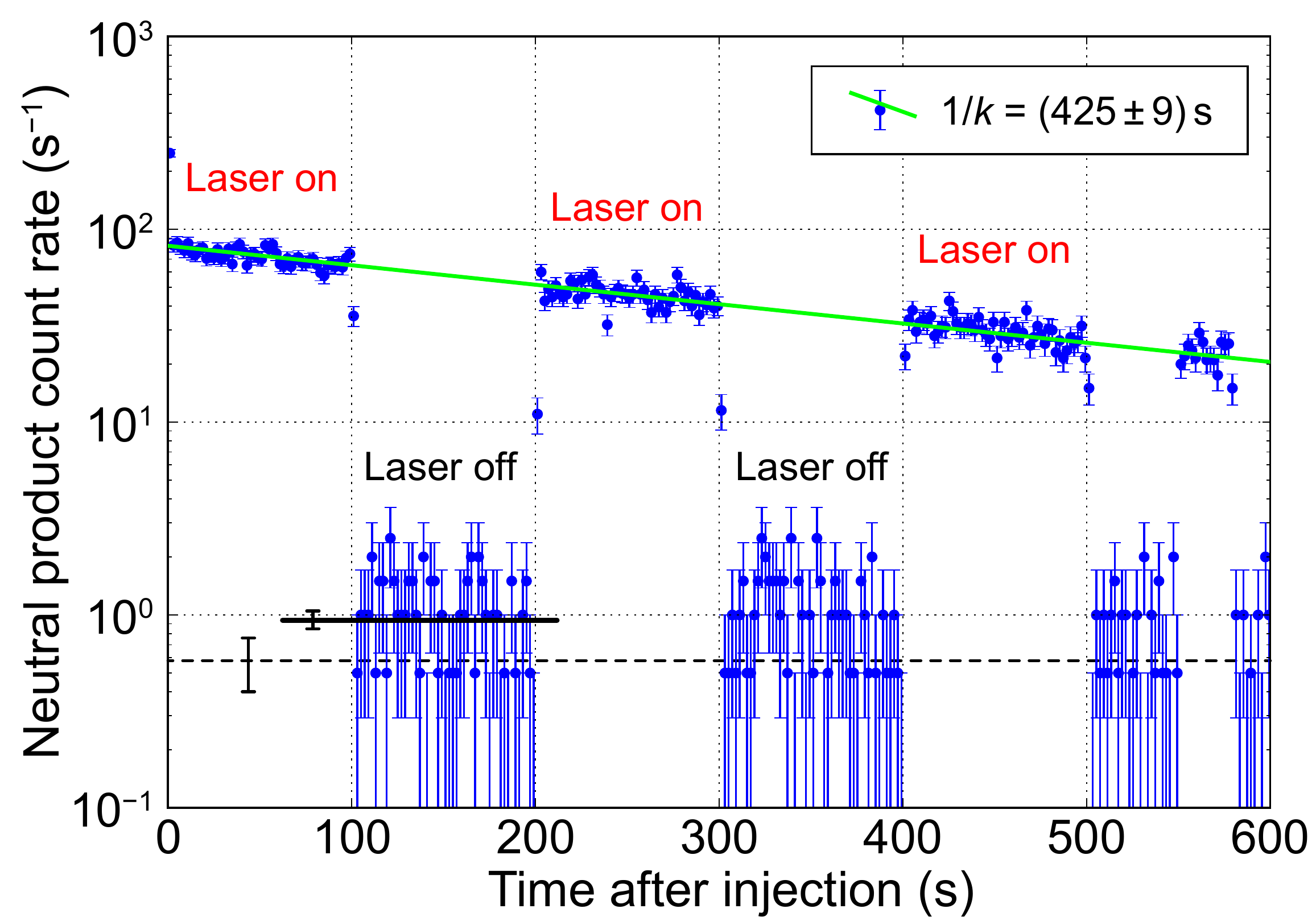}
    \caption{Count rate of neutral products photodetached by a HeNe laser from a 60~keV OH$^-$ beam in the CSR (temperature $\sim$\,6~K) and detected by COMPACT as a function of storage time.  During ``laser on'' the count rates are significantly higher than the dark count rate of the detector.  The fit of an exponential time dependence to these rates yields the given decay time constant.  In ``laser off'' periods the count rate only slightly exceeds the dark count rate.  A full horizontal line indicates the average of the count rate in the first ``laser off'' period and its one-standard-deviation error limits.
The ion beam is dumped at 570~s, with the dark count rate alone showing at later times (average indicated by the dashed horizontal line with errors).}
    \label{fig:lifetime:oh}
\end{figure}

To artificially create a signal proportional to the particle number in the stored ion beam, we stored a beam of hydroxyl anions (OH$^-$) at 60~keV and overlapped the ion beam with a cw HeNe laser operating at a wavelength of 633~nm (see Sec.\ \ref{lasersection}).  The laser photon energy of 1.96~eV was sufficient to photodetach the excess electron in the stored OH$^-$ ions whose electron affinity for ground-state ions \cite{smith_high-resolution_1997} is 1.828~eV.  For constant laser power, the created neutral rate on the detector gives a relative measure of the ion-beam intensity.  The neutral rate observed over 10~min following the ion injection is shown in Fig.~\ref{fig:lifetime:oh}.  The initial number of stored ions for this measurement can be calculated (see Sec.\ \ref{current}) as $N_i=6.8_{-4.6}^{+0.9}\times10^7$.  In the ``laser on'' phases, this measurement yields a beam-decay time constant of $(425\pm9)$~s.  In the ``laser off'' phases, it also illustrates that the detachment rate by ion collisions with residual-gas molecules in the cryopumped inner vacuum system falls to a very low level.  Taking the difference of the average rates in the first ``laser off'' phase and at $>570$~s (no ion beam), we find an ion-induced rate of $(0.37\pm0.21)$~s$^{-1}$ whose significance will be analyzed below.

\begin{figure}
    \centering
    \includegraphics[width=82mm]{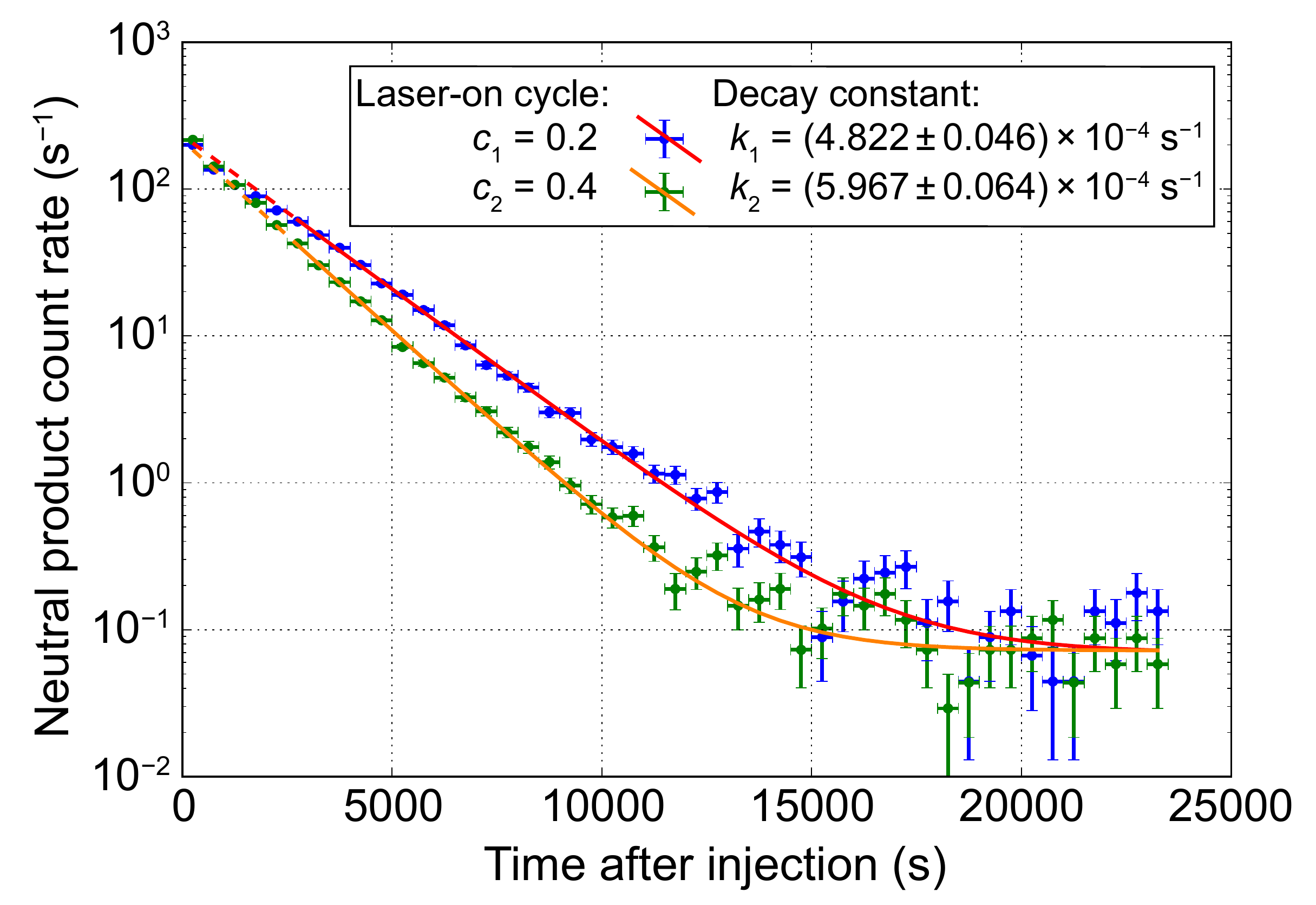}
    \caption{Count rate of neutral products photodetached by a HeNe laser from a 60~keV Ag$_2^-$ beam in the CSR (temperature $\sim$\,6~K) and detected by COMPACT as a function of the storage time $t$. The laser was modulated between ``on'' and ``off'' at a period of 500~s with duty cycles $c_1$ and $c_2$ as given, and the rates during the ``on'' phases are plotted.  Exponential fits of these rates for $t>3000$~s including a common, estimated constant background rate yield the given beam decay constants.}
    \label{fig:lifetime:ag}
\end{figure}

In further measurements we found that the ion-beam lifetime depended on the mass of the stored ions and considerably increased when the 60~keV OH$^-$ anions (mass 17~u) were replaced by heavier species.  The detailed scaling with the ion mass is under study.  The longest lifetime measured in the CSR was obtained using the silver dimer anion Ag$_2^-$ (mass 216~u, photodetachment energy \cite{ho_photoelectron_1990} 1.023~eV).  As for the OH$^-$ beam, the neutral rate arising from residual-gas collisions almost vanished in the detector dark-count rate.  The HeNe laser was used to probe the ion-beam intensity by photodetachment as a function of storage time.  To account for the depletion of the ion beam by the photodetachment, the laser beam was switched between ``on'' and ``off'' phases using two different duty cycles, $c_1$ and $c_2$.  The beam decay constants $k_1$ and $k_2$ measured in these two cases (see Fig.~\ref{fig:lifetime:ag}) were used to calculate the undisturbed decay constant $k_0=(c_2k_1-c_1k_2)/(c_2-c_1)$ of the Ag$_2^-$ anions, yielding $k_0=(3.68\pm0.11)\times10^{-4}$~s$^{-1}$.  The beam storage lifetime (time constant) of Ag$_2^-$ ions undisturbed by laser photodetachment then results in $1/k_0 = (2717\pm81)$~s.  As demonstrated by Fig.~\ref{fig:lifetime:ag}, the long storage lifetime enabled us to find a significant signal of neutral products still more than three hours ($\sim$\,10$^4$~s) after the injection.

We can use this measurement also to verify the fraction of photodetachment events counted by the COMPACT detector.  The laser-induced decay constant is found from the measurement of Fig.~\ref{fig:lifetime:ag} to be $k_L=(k_2-k_1)/(c_2-c_1)=(5.7\pm0.4)\times10^{-4}$~s$^{-1}$.  With the initial stored ion number for this measurement of $N_i=4.4_{-2.6}^{+0.6}\times10^6$ and the observed initial laser-induced count rate $R_{i,m}=200$~s$^{-1}$, this yields a fraction of observed events of $\epsilon=R_{i,m}/k_LN_i=0.09_{-0.01}^{+0.12}$.  This number represents the product of the geometrical and the counting efficiencies.  At room temperature and with particles of higher velocity, a counting efficiency close to unity was demonstrated for the COMPACT detector \cite{Spruck:RSI:2015} for a suitable setting of the pulse-height detection threshold, discriminating the background pulses from those by heavy-particle impact onto the converter cathode.  Measurements in the CSR at the final cryogenic temperature showed that the pulse-height distributions behaved similar, although the discriminator settings had to be turned up substantially in order to suppress an electronic noise background.  With this and with the geometrical losses expected in the product detection scheme we can rationalize the value of $\epsilon$ determined in the described procedure.

The upper limit of the residual-gas induced detachment rate $R_{g,m}$ that can be measured by the detector in the absence of photo-induced processes as shown in Fig.~\ref{fig:lifetime:oh} (OH$^-$) places an upper limit on the residual-gas density in the straight section of the CSR (length $l_b$) aligned with the detector.  In a separate measurement, we could determine the detachment cross section for OH$^-$ ions of 35~keV on H$_2$ as $(7.2\pm0.5)\times10^{-16}$~cm$^2$ with a slight increase of order $0.1\times10^{-16}$~cm$^2$ between 25 and 35~keV.  Hence, we can safely assume for OH$^-$ a residual-gas detachment cross section of $\sigma>6\times10^{-16}$~cm$^2$ and from this constrain the rate of neutral OH products moving toward the detector as $R_g=N_if_0n_gl_b\sigma$, where $n_g$ denotes the residual-gas density.  With the length $l_b=3.0$~m, the revolution frequency $f_0=23.47$~kHz, the lower limit of the efficiency, $\epsilon=R_{g,m}/R_g>0.07$, and the upper limit of the ion-induced count rate from Fig.~\ref{fig:lifetime:oh}, $R_{g,m}<0.6$~s$^{-1}$, the residual-gas density can be constrained to $n_g=R_{g,m}/\epsilon N'_if_0l_b\sigma<140$~cm$^{-3}$.  Here the ion number $N'_i$ at 150~s (average time of the first ``laser off'' window) is assumed to be 0.7 of the initial ion number (i.e., $N'_i>1.54\times10^{7}$ OH$^-$ ions).  The density limit is equivalent to a pressure limit of $p=n_gk_BT<5.8\times10^{-15}$~mbar at room temperature ($T=300$~K).  This indicates that the CSR, with the inner vacuum chamber at $\sim$\,6~K, has reached a room-temperature equivalent pressure level well below 10$^{-14}$~mbar, surpassing the design goal of an equivalent pressure of 10$^{-13}$~mbar, previously set for the CSR, by more than an order of magnitude.

\section{Conclusions and Outlook}
\label{outlook}

The CSR was successfully cooled down to $\sim$\,6~K and has stored anion and cation beams in the 60~keV range with beam-decay time constants up to about an hour.  Non-destructive diagnostic measurements showed that the ion optics are well understood and that various tools are available for tuning the machine to the experimental requirements.  The beam lifetime does not appear to be limited by inelastic collisions of the stored ions in the residual gas.  In fact, the momentum distribution of the stored ions, measured by the Schottky-noise spectrum, is found to broaden when observed over long times in a manner that can be well reproduced assuming that the ion energy slightly changes in a diffusion-type process (Sec.\ \ref{noise}).  Future studies of this effect may reveal the origin of the limitations on the beam lifetimes.  Nevertheless, the observed beam lifetimes are by far long enough for most envisaged experiments in atomic, molecular and cluster physics.  

For future experiments, the ion optical layout of the CSR will allow the beam energy to be increased to 300~keV for singly charged ions.  The ring also offers long field-free straight sections for further experimental equipment.  One of the first instruments to be added is the photocathode-generated merged electron-beam device, which will enable phase-space cooling of stored ion beams.  With envisaged electron-beam energies \cite{shornikov_maximum_2014} down to $\sim$\,1~eV, electron cooling is expected to work for singly charged stored ion beams of a mass up to $\sim$\,165~u at 300~keV.  The detectors for photodetachment products (Sec.\ \ref{detectors}) operated in the present studies will also serve for the neutral products from inelastic collisions of stored ions in the merged electron beam.  A photon--ion interaction zone will continue to be available in a modified layout after installation of the merged electron-beam device.

Installation of additional experimental equipment at the CSR is already foreseen.  A high-power diode-laser interaction region, similar to that realized in recent work \cite{oconnor_generation_2015}, is in preparation for producing a fast beam of neutral atoms merging with the stored ions along a straight section of the CSR (see Sec.~\ref{concept}).  This will be the basis for an experimental program of studying reactive low-energy collisions between atoms and molecular ions in a cryogenic environment.  Furthermore, aiming at the mass-sensitive detection of neutral fragmentation products from the interaction region of the stored ions with the merged electron beam, a multi\-pixel cryogenic microcalorimeter is in preparation.  The high mass resolution of this detection technique for neutral molecular fragmentation products from fast beams has recently been demonstrated \cite{novotny_cryogenic_2015}.  Also the development of ion sources for large molecular and cluster ions for the 300~kV acceleration platform is underway.  The performance of the CSR confirmed by the present study, as well as first measurements published recently \cite{oconnor_photodissociation_2016}, underline that the machine will be a powerful instrument for a broad range of physics experiments with fast beams of atomic, molecular and cluster ions over long observation times in a cryogenic environment and at extremely low residual-gas density.

\begin{acknowledgements}
  The commitment of the technical and engineering staff at the Max Planck Institute for Nuclear Physics in the design and setup of this complex machine is greatly appreciated.  For contributing technical solutions on critical components and sub-systems, special thanks are due to R.~Epking, K.~Hahn, D.~Kaiser, M.~K\"onig, O.~Koschorreck, V.~Mallinger, F.~M\"uller, T.~Spranz, T.~Weber and P.~Werle.  For early discussions on the cryogenic realization we thank B. Petersen (DESY) and C. Schr\"oder (GSI).  We thank H.~Quack and C.~Haberstroh (TU Dresden) and C. Day (KIT) for support in realizing the cryogenic vacuum system and K. Stiebing (Goethe Universit\"at Frankfurt am Main) for support on injection beamline components.  A.~B., K.~S. and S.~S.~K. were supported by the Deutsche Forschungsgemeinschaft (DFG) within the DFG Priority Program 1573 ``Physics of the Interstellar Medium'' (contract numbers WO 1481/2-1, Schi 378/9-1 and KR 4617/1-1, respectively).  F.~G., E.~A.~G., A.~P.~O. and H.~K. were supported by the European Research Council under Grant Agreement No.\ StG~307163.  O.~N. was supported in part by the NSF Division of Astronomical Sciences Astronomy and Astrophysics Grants program and by the NASA Astronomy and Physics Research and Analysis Program.  D.~S. acknowledges support by the Weizmann Institute through the Joseph Meyerhoff program.  X.~U. acknowledges support from the Fund for Scientific Research (FNRS) and the IISN under Contract No.\ 4.4504.10.  The support by the Max Planck F\"orderstiftung is gratefully acknowledged.
\end{acknowledgements}

\bibliography{csrmachine}

\end{document}